\documentclass[fleqn,usenatbib]{mnras}
\usepackage{newtxtext,newtxmath}
\usepackage[T1]{fontenc}
\usepackage{epsfig}
\usepackage{amsmath}
\usepackage{color}
\usepackage{booktabs}
\usepackage{pdflscape}

\usepackage[usenames]{xcolor}
\usepackage{aas_macros}  % required to handle the ADS bibtex journal macros
\usepackage{chngcntr}

\newcommand{\Msolar}{\mbox{\,$\rm M_{\odot}$}}        % solar mass
\newcommand{\Rsolar}{\mbox{\,$\rm R_{\odot}$}}        % solar radius
\newcommand{\Lsolar}{\mbox{\,$\rm L_{\odot}$}}        % solar luminosity
  \newcommand{\Teff}{\mbox{\,\em T$_{\rm eff}$}}         % effective temperature
  \newcommand{\kmsec}{\,\mbox{$\mbox{km}\,\mbox{s}^{-1}$}}    % kilometres/second
  \newcommand{\cmss}{\,\mbox{$\mbox{cm}\,\mbox{s}^{-2}$}}    % centimetres/second/second
  \newcommand{\kelvin}{\,\mbox{K}}                       % K Kelvin
  \def\simge{\mathrel{\raise1.16pt\hbox{$>$}\kern-7.0pt
    \lower3.06pt\hbox{{$\scriptstyle \sim$}}}}           % approx ge
  \def\simle{\mathrel{\raise1.16pt\hbox{$<$}\kern-7.0pt
    \lower3.06pt\hbox{{$\scriptstyle \sim$}}}}           % approx le
\title[V652\,Her: non-linear pulsation models]{Improved hydrodynamic pulsation models for the pulsating extreme helium star V652\,Herculis}
\author[Jeffery, C.~S., et al.]{ C. Simon Jeffery$^{1,2}$, Pilar Monta\~n\'es-Rodr\'iguez$^{1,3,4}$, Hideyuki Saio$^{5}$\\
$^{1}$Armagh Observatory and Planetarium, College Hill, Armagh BT61 9DG, UK\\
$^{2}$School of Physics, Trinity College Dublin, Dublin 2, Ireland\\
$^{3}$Instituto de Astrof\'{\i}sica de Canarias, C/V\'{\i}a L\'actea s/n, E-38200 La Laguna, Spain\\
$^{4}$Departamento de Astrof\'{\i}sica, Universidad de La Laguna, Av., Astrof\'{\i}sico Francisco S\'anchez, s/n, E-38206 La Laguna, Spain\\
$^{5}$Astronomical Institute, School of Science, Tohoku University, Sendai 980-8578, Japan
}

\begin{document}

\date{Accepted \ldots. Received \ldots; in original form \ldots}

\pagerange{\pageref{firstpage}--\pageref{lastpage}} \pubyear{2021}

\maketitle

\label{firstpage}

\begin{abstract}
New non-linear hydrodynamic models have been constructed to simulate the radial pulsations observed in the extreme helium star V652\,Her. These use a finer zoning to allow higher radial resolution than in previous simulations. 
Models incorporate updated OPAL and OP opacity tables and adopt a composition based on the best atmospheric analyses to date. 
Key pulsation properties including period, velocity amplitude and shock acceleration are examined as a function of the mean stellar parameters (mass, luminosity, and effective  temperature).  
The new models confirm that, for large amplitude pulsations, a strong shock develops at minimum radius, and is associated with a large 
phase delay between maximum brightness and minimum radius. 
Using the observed pulsation period to constrain parameter space in one dimension, other pulsation properties are used to constrain the model space further, and to critically discuss observational measurements.
Similar models may be useful for the interpretation of other blue large amplitude pulsators, which may also exhibit pulsation-driven shocks. 
\end{abstract}

\begin{keywords}
             stars: chemically peculiar,
             stars: individual (V652 Her),
             stars: oscillations,
             shock waves
\end{keywords}

\section{Introduction}
\label{s:intro}
V652 Her is a unique object for several reasons:
a) it is an extreme helium star, belonging to a group of approximately 15 low-mass giants with effective temperatures in the range 10\,000 to 30\,000 K and  very low surface hydrogen abundances \citep{berger63.v652}, 
b) it shows large-amplitude radial pulsations which have allowed direct measurement of its radius and the
 possibility to explore the physical properties of its interior \citep{landolt75,hill81,lynasgray84,jeffery01b}, 
c) lying outside the classical instability strip, the pulsations could only be explained by the introduction of additional opacity from iron-group elements \citep{saio93},
d) the pulsation period is decreasing, which implies a secular radius contraction \citep{kilkenny84}, and    
e) it has a nitrogen-rich, carbon-poor surface which cannot be easily explained by a single-star evolution model \citep{jeffery99,przybilla05}.
A comprehensive review of these properties was most recently given by \citet{jeffery15b}; the most precise measurements to date give a mean radius $\langle R\rangle=2.31\pm0.02\Rsolar$, effective temperature $\langle T_{\rm eff}\rangle=20\,950\pm70 \kelvin$,  luminosity $\langle L\rangle=919\pm14\Lsolar$ and mass  $M=0.59\pm0.18\Msolar$ \citep{jeffery01b}. 
A summary is shown in Table\,\ref{t:pars}

Such a combination of features makes this peculiar object extremely interesting and 
presents challenges for theories of stellar evolution and pulsation. 
Models of the post-merger evolution  of  two helium white dwarfs match almost all of its observational properties \citep{saio00,zhang12a}. 
Other models, such as the evolution of a post-giant branch star following a very-late core helium flash \citep{brown01,byrne18a} cannot be ruled out. 

Hydrodynamic models of the pulsations in V652\,Her have been calculated by \citet{fadeyev96} and \citet{montanes02}. 
These provided the first such calculations using opacities enhanced by contributions from iron-group elements, using OP and OPAL opacities respectively \citep{Seatonetal1994,iglesias96}; \citet{montanes02} also provided models for the similar pulsating helium star BX\,Cir. 

Meanwhile an extensive new spectroscopic study of the pulsation has defined the radial velocity curve very precisely,  mapping the motion of the mean surface, and providing a depth-dependent description of motion  as a function of optical depth within the photosphere \citep{jeffery15b}.
These observations indicate  that the photosphere is compressed by a factor $\approx2$ at minimum radius, and show that the phase of minimum radius is a function of optical depth, travelling through the photosphere at a speed between 140 and 240 \kmsec. 
The  acceleration at minimum radius is largest very deep in the  photosphere, where individual line profiles imply jump discontinuities of over 70\kmsec\ in 150\,s and provide evidence of a pulsation-driven shock. 
The fundamental  observable remains the  amplitude of the surface velocity. 
This depends marginally on which point of the photosphere is being measured, and also on the transformation of this quantity from the observers frame to the stellar rest frame.
With an observed amplitude of 68\kmsec\, and a projection factor $p\approx 1.31$ \citep{montanes01}, the un-projected amplitude obtained from the strong N{\sc ii}4621\AA\  line is $\Delta u=89\kmsec$.  

In order to obtain higher resolution in the hydrodynamic calculations and to address the questions raised by the latest observations, we set out to compute a new generation of hydrodynamic radial pulsation models for V652\,Her. 
The aim of these is to reproduce the observed period and amplitude, and also the detailed features of the velocity curve.   
In doing so, we  investigate how much the fundamental properties of the star are constrained by the pulsation models,  and by how much these are at variance with measurements of these properties obtained by other means. 
We  investigate if and how the pulsation properties vary with period, since the latter has decreased by $\approx1\%$ since discovery.   
The models are also required in order to construct dynamical model atmospheres necessary for the accurate interpretation  of the observed spectrum and hence to deduce more reliably the mean atmospheric properties.

\begin{table}
\caption{Published surface, pulsation and other properties of V652\,Her. 
Notes. 1:\citet{kilkenny05}, 2:\citet{jeffery01b}, 3:\citet{przybilla05}, 4:\citet{jeffery99}, 5:\citet{jeffery15b}.  }
\label{t:pars}
\begin{tabular}{l rr ll}
\hline
$P_0$ (1974) & \multicolumn{2}{c}{$0.107\,992\,82\pm0.000\,000\,02$}   & d & 1 \\
$\dot{\Pi}/\Pi$ & \multicolumn{2}{c}{$-7.426\pm0.005\times10^{-8}$} & d\,d$^{-1}$ & 1 \\[2mm]
$\langle T_{\rm eff} \rangle$ (ion equil) & 22\,930 & $\pm 10$ & K & 2 \\
$\langle \log g \rangle$ (He {\sc i} lines) & 3.46 & $\pm 0.05$ & cm\,s$^{-2}$ & 2 \\
$\langle T_{\rm eff} \rangle$ (total flux) & 20\,950 & $\pm 70$ & K & 2 \\
$\langle R \rangle$  & 2.31 & $\pm 0.02$ & \Rsolar & 2 \\
$\langle L \rangle$ & 919 & $\pm 014$ & \Lsolar & 2 \\
$M$ & 0.59 & $\pm 0.18$ & \Msolar & 2 \\
$d$ & 1.70 & $\pm 0.02$ & kpc & 2 \\[2mm]
$T_{\rm eff}$ (NLTE) & 22\,000 & $\pm 500$ & K & 3 \\
$\log g $ (NLTE) & 3.20 & $\pm 0.10$ & cm\,s$^{-2}$ & 3 \\
$n_{\rm H}$ (NLTE ) & 0.005 & $\pm 0.0005$ &   & 3 \\[2mm]
$T_{\rm eff}$ (mean) & 24\,550 & $\pm 500$ & K & 4 \\
$\log g $ (mean) & 3.68 & $\pm 0.10$ & cm\,s$^{-2}$ & 4 \\
$\log n_{\rm H}$  & --2.16 & $\pm 0.07$ &  & 4 \\
$\log n_{\rm He}$  & 0 &      &  & 4 \\
$\log n_{\rm C}$  & --4.40 & $\pm 0.27$ &  & 4 \\
$\log n_{\rm N}$  & --2.61 & $\pm 0.06$ &  & 4 \\
$\log n_{\rm O}$  & --4.00 & $\pm 0.08$ &  & 4 \\
$\log n_{\rm Fe}$  & --4.14 & $\pm 0.15$ &  & 4 \\[2mm]
$\Delta v$ & 68 & & \kmsec & 5 \\
$\Delta u = p \Delta v$ & 89 & & \kmsec & 5$^{\dagger}$ \\
$\Delta \phi$ & +0.13 &$\pm 0.01$ & cycles & 5$\ddagger$ \\
$T_{\rm eff}$ ($R_{\rm max}$) & 22\,500 & -- & K & 5$^{\ast}$ \\
$\log g $ ($R_{\rm max}$) & 3.3 & -- & cm\,s$^{-2}$ & 5$^{\ast}$ \\[2mm]

\hline
\multicolumn{5}{l}{$\dagger$: projection factor $p = 1.31$: \citet{montanes01}} \\
\multicolumn{5}{l}{$\ast$: adopted model}\\
\multicolumn{5}{l}{$\ddagger$: light maximum defined by optical light curve: \citet{kilkenny05}}\\
\end{tabular}
\end{table}

\begin{figure}
\centering
\includegraphics[width=0.98\linewidth]{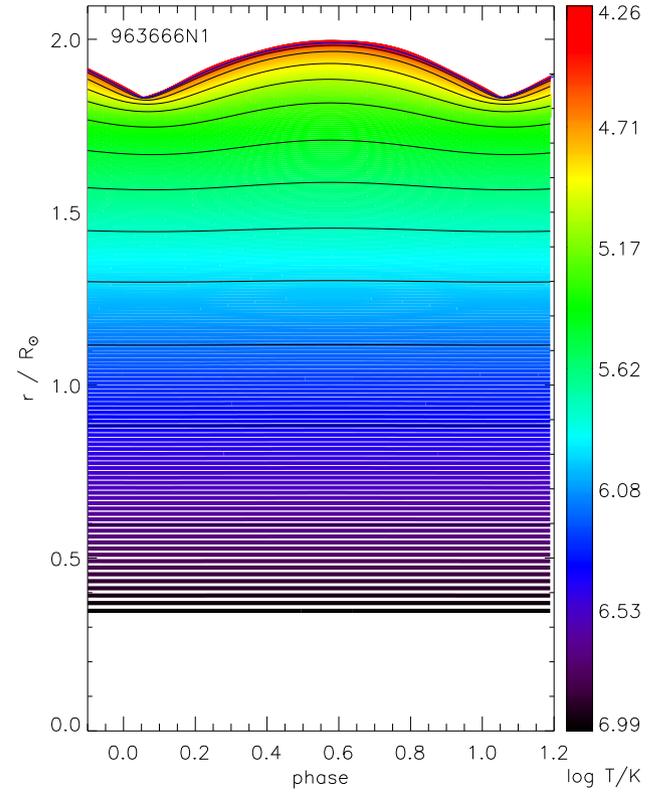}
\caption{The position $r$ of mass zones in the model interior throughout a pulsation cycle. This figure shows model 963666N1, having $M = 0.66\Msolar$, $\log L/\Lsolar = 2.96$,  $\log \Teff/{\rm K} = 4.36$ and mixture N1 (see text). 
The mass zones are colour-coded for temperature. 
The base of the photosphere, defined by Rosseland mean optical depth $\tau_{\rm Ross}=2/3$, is shown by a bold dark line. The position of every 20th zone is marked by a thin black line. Fig.\,\ref{f:surf} shows parts of this model in more detail. 
}
\label{f:guts}
\end{figure}

The computation assumptions, code and parameter space are described in \S\,\ref{s:hydro}.  Results for pulsation periods, velocity amplitudes and curve shapes
are presented in \S\,\ref{s:periods}. Internal properties of the pulsation are explored and compared with the observed properties of V652\,Her in \S\,\ref{s:match}. \S\,\ref{s:conc} concludes.      

%%====================================================================

\section[]{Hydrodynamic calculations}
\label{s:hydro}

The models were computed with the \cite{montanes02} non-linear radial pulsation code, which is based on 
a prescription by \cite{christy67} and an implementation by \cite{bridger83}.

The Lagrangian coordinate is used, in which the mass within each zone $\delta m_{i-1/2} = m_i -m_{i-1}$. We define $i$ to increase outward and $\delta m_{i-1/2} = \alpha\delta m_{i+1/2}$ with $\alpha$ constant and $\lesssim 1.1$ to yield a logarithmic spacing in mass \citep[cf.][]{christy67}.

The stellar core is not computed.
Instead, the mean energy flux from the core is assumed constant, with no nuclear energy released in the envelope.  
With this approximation, the radial pulsation is described through the momentum equation and the heat flow equation. 
Radiation diffusion is regarded as the only energy transport mechanism. 
Convective energy transport is assumed to be negligible, which is reasonable for the envelopes of stars with effective temperatures greater than 20\,000\,K. 
This provides four first-order differential equations in radius $r$, energy flux $l$, density $\rho$ and temperature $T$ in 
 terms of mass $m$ and time $t$.

Pressure $P$, electron density $n_{\rm e}$, internal energy $E$ and their first derivatives are computed as functions of temperature and density using an equation of state (EOS) \citep{bridger83} based on the formalism given by \citep[chapter 5]{mihalis78}. 
Opacity $\kappa$ and its derivatives are interpolated as a function of temperature and density from the OPAL95 opacity tables \citep{iglesias96}. 

The Bridger EOS solves the Saha ionization equation for 4 species, hydrogen, helium, nitrogen and magnesium, with the latter representing 80\% and 20\% by mass of the metals ($Z$), respectively. It treats ions up to the 6th state ($5+$) and is therefore valid up to $\log T \approx 5.3$. Above this temperature it starts to underestimates the ionisation fraction, leading to a maximum error of 0.3\% in the mean mass per particle when full ionization is achieved (at $\approx 10^7$ K), and hence in quantities which depend on this value 
More sophisticated EOS required for opacity calculations \citep[e.g.][et seq.]{hummer88} have minor influence on the values of major state variables in hydrogen or helium-rich mixtures except at electron-degenerate densities \citep{rogers96}. Molecules do not feature in V652\,Her. 

Four boundary conditions are required. 
At the surface, we require i) the gas pressure to be zero and ii) the usual relation between radius, luminosity and effective temperature. 
At the inner boundary, defined where the radius $R/10< r < R/5$  or the  temperature $T = 10^8\,\kelvin$, the dynamical model 
should be stable and therefore iii) $\dot{\rho}_{1}=0$, $\dot{r}_1\equiv u_1 =0$, $\dot{l}_1=0$  and $\dot{T}_1=0$ at the inner boundary.   

In the non-linear treatment, the complete system of differential equations can be written in a finite-difference form and 
can be solved as an initial value problem. 
As initial value, we compute a finely-zoned envelope in hydrostatic 
equilibrium; this is subsequently rezoned onto a Lagrangian grid with $n$ mass zones. 
A small velocity perturbation
\[
u^1_{i} = 0.1 \left(\frac{r^1_{i}}{r^1_{n}}\right)^4
 \kmsec
\]
is applied to the initial envelope model, where $r^1_i$ and $r^1_n$ are the initial radii of the $i$th and $n$th zones, respectively. 
Then we follow the growth of the perturbation until a stable pulsation period is obtained. 

\begin{figure*}
\centering
\includegraphics[width=0.40\linewidth]{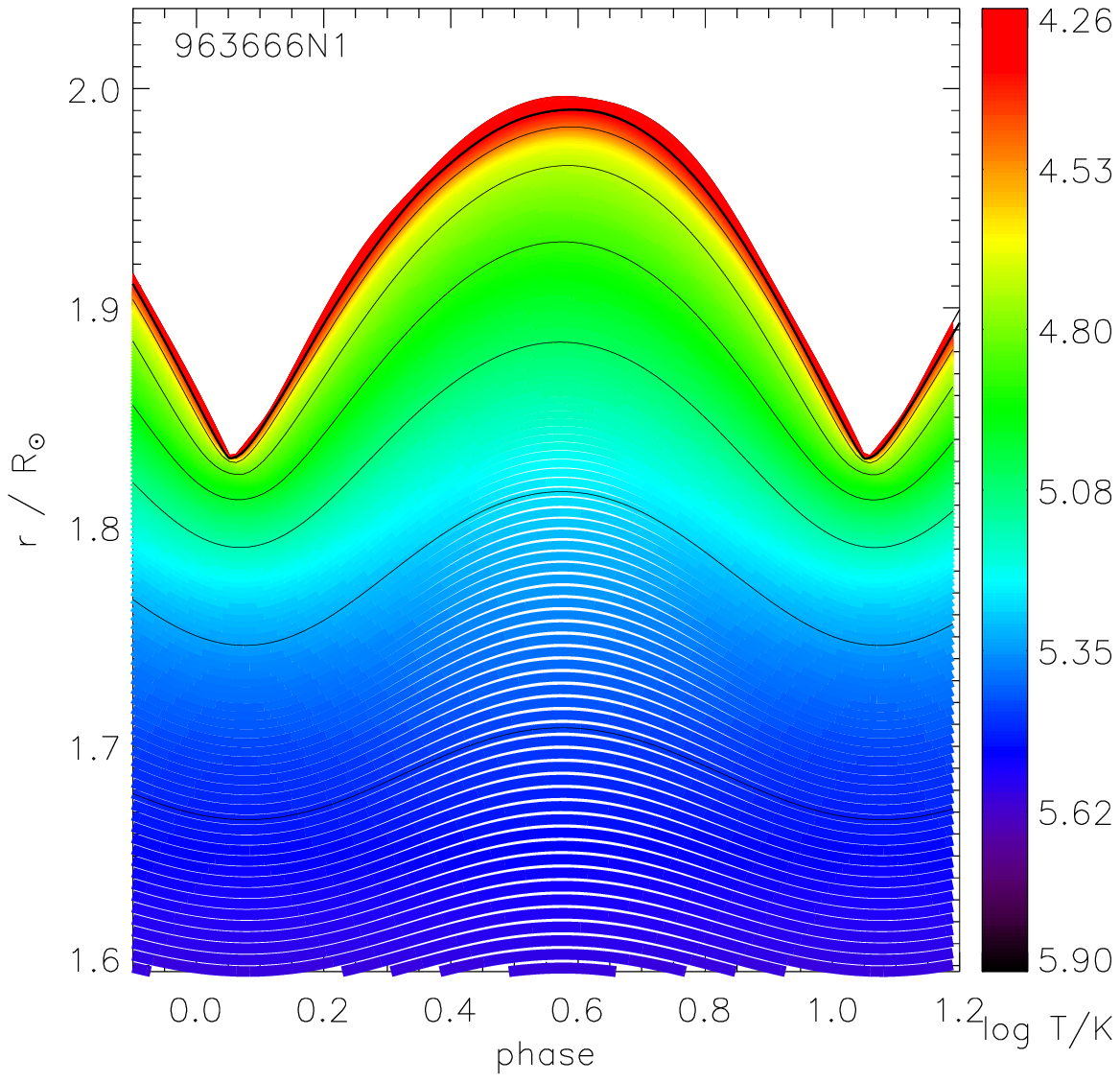}
\includegraphics[width=0.40\linewidth]{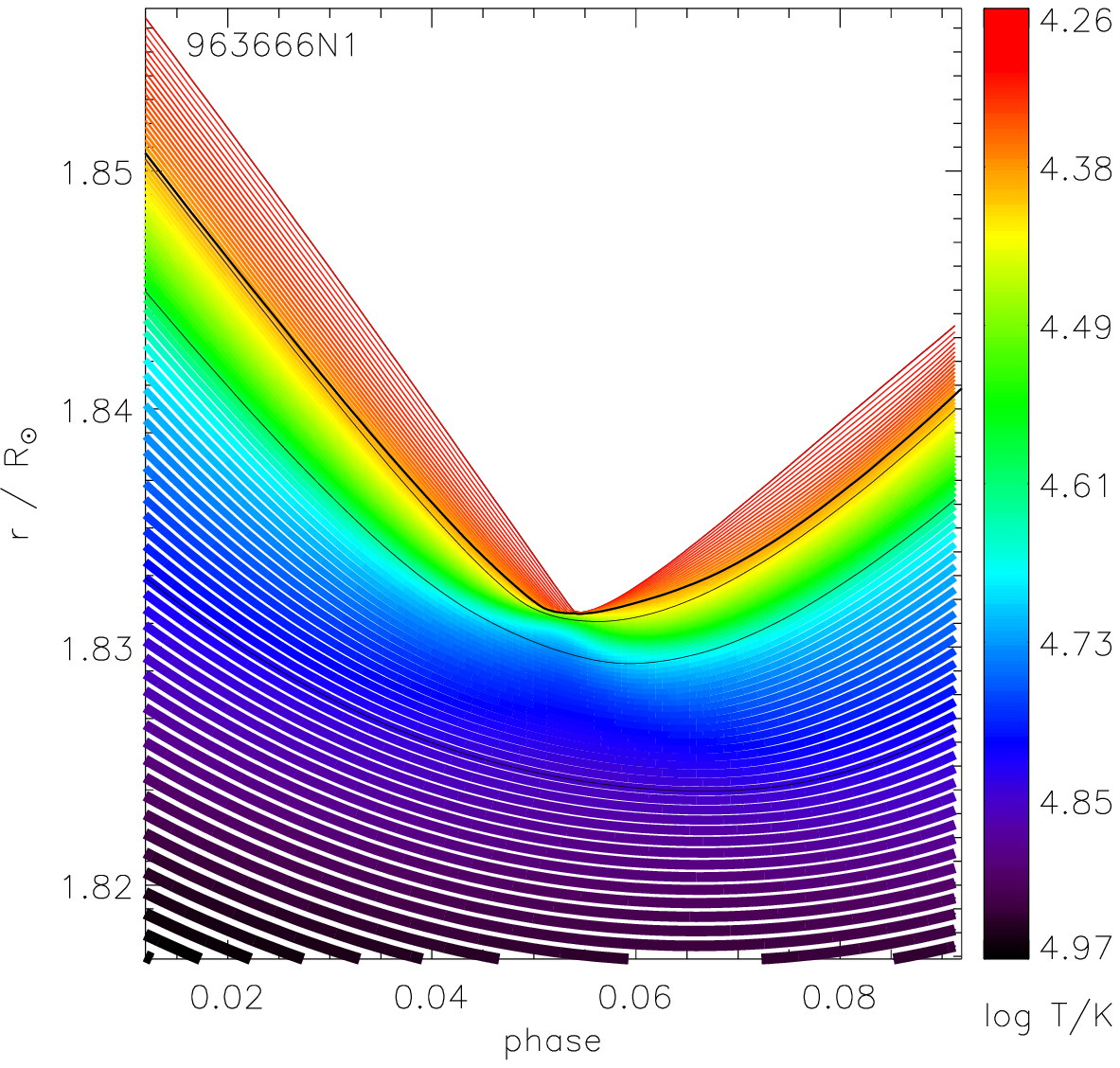}
\caption{As Fig.\,\ref{f:guts} for (left) outer layers including the opacity driving zone and (right) layers close to the surface during passage through minimum radius . 
}
\label{f:surf}
\end{figure*}

A limit cycle is deemed to have been reached when strict periodicity is achieved. 
We follow the subsequent temporal evolution step by step until pulsations  with nearly constant amplitude and period are achieved (successive cycles differ by less than 0.5\%). 
This is considered to be the solution for the initial value  problem, and constitutes a hydrodynamic description of the pulsation through the stellar envelope (Fig.\,\ref{f:guts}). 
In the current study, between 4\,000 and 10\,000 
 pulsation cycles for stellar envelopes with 200--250 mass zones were computed  to reach this final level of convergence, in contrast to the 200--500 cycles for envelopes with 40--50 mass zones computed in the previous models \citep{montanes02}.

An artificial viscosity dissipation following \citet{stellingwerf75} was used to deal with the shocks produced in outer layers during a supersonic contraction and bounce. 
This is especially important around minimum radius, as shown in Fig.\,\ref{f:surf}.  
The artificial viscosity, which is proportional to the square of compression rate, is included where matter is compressed  to prevent sharp discontinuities appearing in the physical variables between adjacent zones.
The latter can lead to numerical singularities causing the calculation to fail. 
\citet{stellingwerf75} introduced a cut-off compression rate, below which the artificial viscosity is set to be zero. This cut-off prevents nonphysical dissipation in slowly contracting layers.
We adopted default values $C_Q=2.0$ and $\alpha_v = 0.1 c_s$ for the constant of proportionality and the cut-off compression rate, respectively.
The adopted $C_Q$ is relatively low \citep{grott05}, but convergence was achieved for nearly all models of interest.

%%====================================================================

\begin{figure}
\begin{center}
\epsfig{file=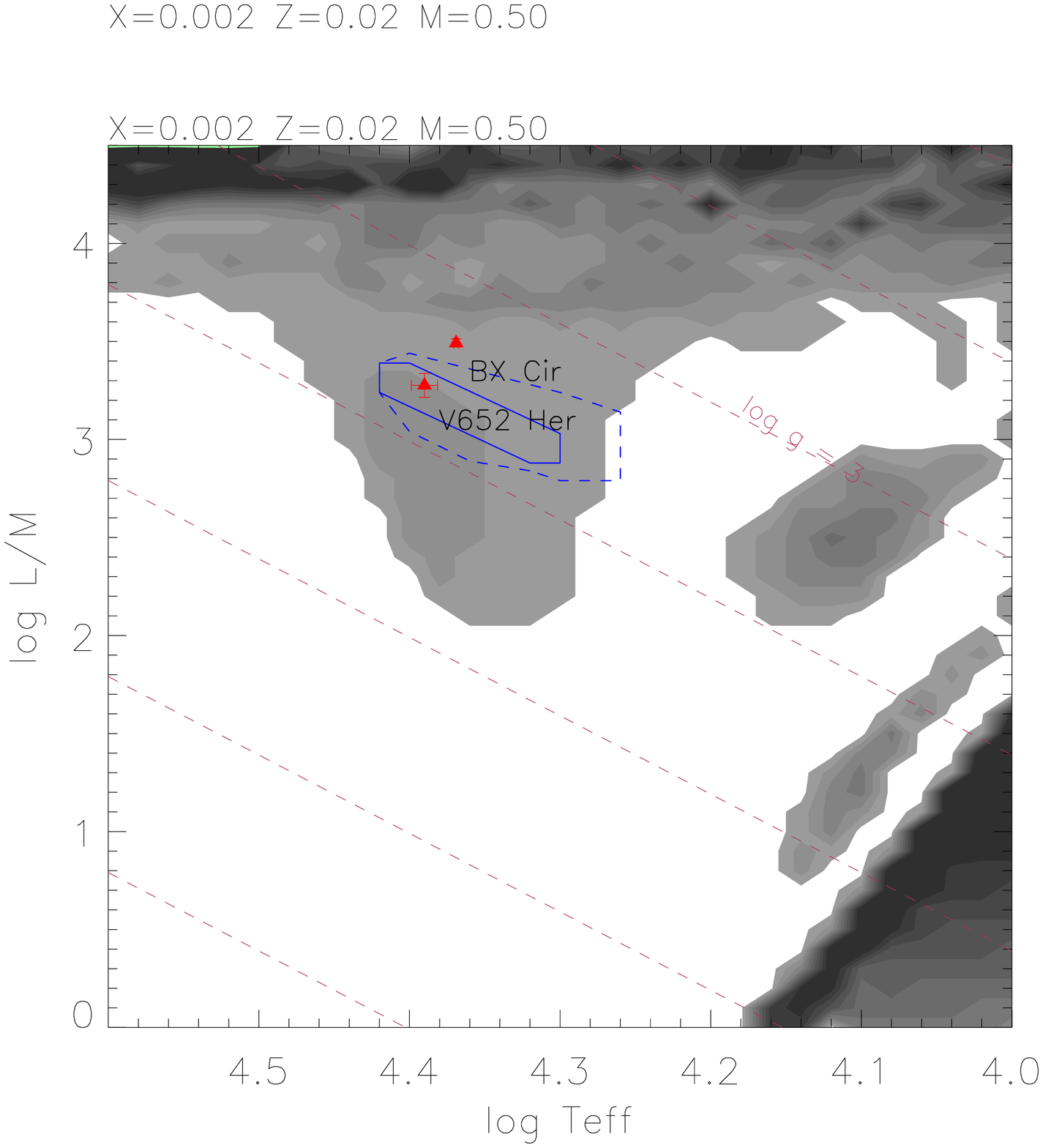,width=8.6cm,angle=0}
\caption[Enlargement]{Unstable pulsation modes in stars with homogeneous
envelopes for composition $X=0.002, Z=0.02$, $M=0.50\Msolar$.
The {\it number} of unstable radial modes in the {\it linear non-adiabatic approximation} is represented by grey scale contours, with the lightest shade marking the instability boundary
(one unstable mode), and the darkest shade representing ten or more more unstable modes. 
 Broken maroon  diagonal lines represent contours  of constant surface gravity at  $\log g = 8, 7, 6, \ldots, 1$. 
The spectroscopic positions of pulsating extreme helium stars V652\,Her and BX\,Cir are marked with red triangles.
The fine grid of {\it non-linear} pulsation models is indicated by a solid blue border; the boundary of the complete grid is indicated by a dashed blue border. The diagram is only weakly mass dependent. 
Adapted from \citet{jeffery16a}. 
}  
\label{f:linear}
\end{center}
\end{figure}

\begin{table}
\caption{Illustration of model grid for $M = 0.66\Msolar$ and mix N1 showing pulsation periods derived from the models.}
\label{t:grid}
\setlength{\tabcolsep}{6pt}
\begin{tabular}{l|cccccc}\hline
& \multicolumn{4}{l}{ $\log \Teff/{\rm K}$} \\
   & 4.30 & 4.32 & 4.34 & 4.36 & 4.38 & 4.40 \\ \cline{2-7}
$\log L/\Lsolar$ & $\Pi$/d  & & & & \\ \cline{2-2}
3.16   &  & & & 0.165 & & 0.121 \\
3.06    & 0.216 & 0.183 & 0.156 & 0.133	& 0.114	& 0.098 \\ 
3.01	&  &  &  & 	0.117 & 0.105 & 0.086 \\
2.96	& 0.175 & 0.150 & 0.127	& 0.108	& 0.094	& 0.083 \\
2.91	&  &  & 0.117	& 0.101 & 0.086 & \\
2.86	& 0.144	& 0.122	& 0.105	& 0.090	& 0.079	& 0.069  \\
2.81	& 0.130 & 0.113	& 0.097	& 0.083 & 0.073 & \\
2.76	& 0.119 & 0.102	& 0.089	& 0.077 & & \\
2.71    & 0.108	& 0.093	& 0.081	& 0.070 & & \\
2.66.   & 0.085 & \\
\hline
\end{tabular}
\end{table}

\subsection{Model grid}
\label{s:grid}

We have set out to identify a model which most closely approximates the observed properties of V652\,Her as outlined in Table\,\ref{t:pars}. 
The best constrained observable is the pulsation period $\Pi$ which is correlated with the square root of the mean density $\langle{\rho}\rangle \propto  \sqrt{ (M/R^3)}$. 
Although the radius has been measured directly from observation, a larger error in the mass arises because the surface gravity $g \propto M/R^2$ is measured spectroscopically, with a fitting error of $\pm0.10$ dex, and a less well determined systematic error. 

With these constraints, we have chosen combinations of \Teff\ and $L$ which yield approximately the correct values of $R$ and $\Pi$, but which also lie on a rectangular grid, i.e. $\log \Teff/{\rm K}\in\{4.30(0.02)4.40\}$ and $\log L/\Lsolar \in\{2.71(0.05)3.06\}$. 
Those  combinations close to the leading diagonal  and having $\Pi\approx0.11\pm0.01$\,d were investigated in greatest detail.
Some outlying models were computed in order to explore the behaviour of amplitude towards the boundaries of the instability zone.  
Fig.\,\ref{f:linear} shows the grid boundaries  together with the instability boundary obtained from a linear non-adiabatic calculation \citep{jeffery16a}. The pulsation topology of the $L/M - \Teff$ plane is relatively insensitive to $M$ \citep{jeffery16a}.

In addition to these, the chemical composition of the stellar envelope (\S\,\ref{s:comp}) represents a free parameter which can be partially constrained by observation. 
This guides the choice of opacity (\S\,\ref{s:opacity}). 
Other choices include the mass of the stellar envelope $M_{\rm env}$ included in the calculation. 
The latter is a computational and not a physical parameter. 

For convenience and without ambiguity, each model is given a label in the form $ llttmmNn $ where 
$ll = {\rm frac}(\log L/\Lsolar)$, 
$tt = {\rm frac}(\log \Teff/{\rm K})$, 
$mm = {\rm frac}(M/\Msolar)$, 
$Nn$ is a letter and integer combination identifying the chemical composition, the choice of opacity table and other properties of the  calculation.    
For example, the label 963666N1 represents a model with 
$\log L/\Lsolar = 2.96$, $\log \Teff/{\rm K}=4.36$, $M = 0.66\Msolar$, and mixture N1$\equiv$J01 (Table\,\ref{t:opacity}). 
For clarity: $\log L/\Lsolar =3.ll$ if $ll \leq 16$ and $\log L/\Lsolar=2.ll$ if $ll \geq 66$.

\begin{table*}
\caption{Normalized relative metal abundances for V652\,Her given by mass and based on \citet{jeffery99} ({\bf J99}) and\citet{jeffery01b} ({\bf J01}). 
Elements marked $\star$ are measured; others are scaled relative to iron. 
{\bf J01} is the metal abundance distribution adopted for the OPAL opacity tables in grid {\bf N1}. A few test models were  calculated with relative abundances as shown in columns {\bf FeNi2} (OPAL)  and {\bf FeNi3}. Abundances relative to {\bf N1} are shown in columns 6 and 8. Test models calculated with mixture {\bf J01} and OP opacities were labelled {\bf N5}.  } 
	\label{t:opacity}
\begin{tabular}{|l|l|l|l|l|l|l|l|}
\hline
Mix & \textbf{J99} &\textbf{J01}     & \textbf{J99/J01} & \textbf{FeNi2}  & \textbf{FeNi2/J01} & \textbf{FeNi3} & \textbf{FeNi3/J01} \\
Label &         & \textbf{N1}  &  & \textbf{N3}  &     & \textbf{N6} & \\
 \hline
\textbf{C}$\star$   & 0.01066 & 0.02559 & 0.4166 & 0.024884  & 0.9724 & 0.025163 & 0.983314 \\ 
\textbf{N}$\star$   & 0.65740 & 0.64285 & 1.0226 & 0.633460  & 0.9854 & 0.618150 & 0.961586 \\
\textbf{O}$\star$   & 0.02678 & 0.04990 & 0.5367 & 0.049171  & 0.9854 & 0.050205 & 1.00613  \\ 
\textbf{Ne}$\star$  & 0.18528 & 0.11698 & 1.5839 & 0.115271  & 0.9854 & 0.112218 & 0.959300 \\
\textbf{Na}  & 0.00107 & 0.00079 & 1.3544 & 0.000778  & 0.9848 & 0.001466 & 1.85570  \\ 
\textbf{Mg}$\star$  & 0.04445 & 0.08282 & 0.5367 & 0.081610  & 0.9854 & 0.079569 & 0.960758 \\ 
\textbf{Al}$\star$  & 0.00239 & 0.00173 & 1.3815 & 0.001705  & 0.9855 & 0.001588 & 0.917919 \\ 
\textbf{Si}$\star$  & 0.02387 & 0.03874 & 0.6162 & 0.038174  & 0.9854 & 0.039879 & 1.02940  \\ 
\textbf{P}$\star$   & 0.00017 & 0.00345 & 0.0493 & 0.003400  & 0.9855 & 0.003168 & 0.918261 \\
\textbf{S}$\star$   & 0.02127 & 0.01473 & 1.4440 & 0.014515  & 0.9854 & 0.015876 & 1.07780  \\ 
\textbf{Cl}  & 0.00019 & 0.00014 & 1.3571 & 0.000138  & 0.9857 & 0.000217 & 1.55000  \\
\textbf{Ar}$\star$  & 0.00415 & 0.00573 & 0.7243 & 0.005646  & 0.9853 & 0.005021 & 0.876265 \\ 
\textbf{K}   & 0.00007 & 0.00005 & 1.4000 & 0.000049  & 0.9800 & 0.000091 & 1.82000  \\ 
\textbf{Ca}  & 0.00134 & 0.00100 & 1.3400 & 0.000985  & 0.9850 & 0.001572 & 1.57200  \\
\textbf{Ti}  & 0.00005 & 0.00004 & 1.2500 & 0.000039  & 0.9750 & 0.000075 & 1.87500  \\ 
\textbf{Cr}  & 0.00027 & 0.00020 & 1.3500 & 0.000197  & 0.9850 & 0.000326 & 1.63000  \\ 
\textbf{Mn}  & 0.00017 & 0.00012 & 1.4167 & 0.000118  & 0.9833 & 0.000168 & 1.40000  \\
\textbf{Fe}$\star$  & 0.01940 & 0.01439 & 1.3482 & 0.028360  & 1.9708 & 0.042991 & 2.98756  \\ 
\textbf{Ni}  & 0.00102 & 0.00076 & 1.3421 & 0.001498  & 1.9710 & 0.002257 & 2.96974  \\ 
\hline
\end{tabular}
\end{table*}

\subsection{Chemical composition}
\label{s:comp}

The choice of initial hydrostatic model makes assumptions about the internal structure and previous evolution of the star. 
For the current calculations, the envelope is {\it assumed} to be chemically homogeneous and free of  energy sources. 
The first leads us to adopt the chemical composition from the surface abundance analyses by \citet{jeffery99,jeffery01b,przybilla05} (Table\,\ref{t:opacity}). 
Potential conditions which might compromise these assumptions are: 
(i) the adopted surface composition was based on static model atmospheres and a mean spectrum around maximum radius; one consequence of continuing work will enable an abundance analysis based on dynamical model atmospheres,  
(ii) apart from iron, the abundances of elements with atomic number $Z>20$ have not yet been measured for V652\,Her,
(iii) diffusion-induced chemical stratification in the stellar interior might already have commenced, particularly in regions of high specific opacity \citep{byrne18a}; however this will be disrupted by pulsations and occasionally by helium shell flashes, 
(iv)  the secular decrease in pulsation period implies that the stellar envelope is contracting and releasing gravitational potential energy as heat; the ratio between the pulsation period and secular period change, or $\dot{\Pi}/\Pi$ is so small as likely to have little effect on the  current hydrodynamic models.

In the case of V652\,Her, various internal structures have been proposed \citep{jeffery84b,saio00}, but have the common feature of a hydrogen-contaminated but predominantly helium-rich envelope surrounding a helium-burning core. 
In the case of the post-merger model, nuclear helium-burning occurs in a thin shell surrounding a degenerate helium core; the radius of the 
latter can be no more than a few $\times10^3$ km, compared with the 1.6 $\times10^6$ km radius of the star. 
In this regard it is therefore safe to integrate down to $r\approx 0.05 R$ before the energy equation becomes an issue (at $< 0.01 R$).

One question concerns the chemical homogeneity of the envelope. 
In  post-merger models, the hydrogen  originates in the residual hydrogen envelopes of the pre-merger helium white dwarfs. 
The merger episode  completely disrupts the entire donor white dwarf, which is then mixed with a fraction of the envelope of its more massive companion, combining some unprocessed hydrogen with CNO-processed helium.  
The composite merger model involves prompt accretion of part of the donor, followed by slow accretion of the remainder from a high entropy "shell". \citet{zhang12a} showed that flash-driven convection following the first helium shell ignition will mix the post-merger envelope, adding some carbon,  but this is then overlaid by nitrogen-rich material accreted from the "shell". 
Whether carbon is  subsequently mixed through the star depends on the total mass and whether the flash-mixed and surface convection zones interact \citep{zhang12a}. 
In low-mass cases, mixing does not occur and the surface remains nitrogen-rich, as in V652\,Her. 
For the 0.25+0.25\Msolar\ composite merger model computed by \citet{zhang12a}, 
the outer 40\% of the mass remains nearly homogeneous; only the first four helium-shell flashes inject a small amount of carbon into the bottom 0.07\Msolar\ of the helium-rich envelope (op cit., Fig. 14). 
In all current calculations, the envelope mass $M_{\rm env} < 0.2\Msolar$ and satisfies the above constraint. 
It remains to be determined whether chemical diffusion driven by radiative levitation could  significantly influence the chemical structure in V652\,Her; we touch on this briefly in \S\,\ref{s:conc}.  

\begin{table}
\caption{The primary grid of completed models with mixture N1 defined in Table\,\ref{t:opacity}. Labels $llttmm{\rm N}n $ are defined in the text.  }
\label{t:models}
\setlength{\tabcolsep}{3pt}
\begin{tabular}{llllll}
\hline
Mass & & & & & \\
\multicolumn{6}{c}{Models} \\
\hline
0.56\Msolar & & & & & \\
 & & & & 063856N1  & 064056N1  \\
 & & & & 013856N1 & 014056N1 \\ 
 & & & 963656N1 & 963856N1 & 964056N1 \\
 & &  913456N1 & 913656N1 & 913856N1 & \\
 & 863256N1 & 863456N1  & 863656N1  & 863856N1 & \\
 & 813256N1  & 813456N1  & 813656N1  & & \\
763056N1  & 763256N1  & 763456N1  & & & \\
713056N1 &  713256N1  & & & & \\[2mm]
 0.66\Msolar & & & & & \\
063066N1 & 063266N1 & 063466N1 & 063666N1 & 063866N1 & 064066N1 \\
         &          &          & 013666N1 & 013866N1 & 014066N1 \\
963066N1 & 963266N1 & 963466N1 & 963666N1 & 963866N1 & 964066N1 \\
         &          & 913466N1 & 913666N1 & 913866N1 &  \\
863066N1 & 863266N1 & 863466N1 & 863666N1 & 863866N1 & 864066N1  \\
813066N1 & 813266N1 & 813466N1 & 813666N1 & 813866N1 & \\
763066N1 & 763266N1 & 763466N1 & 763666N1 & & \\
713066N1 & 713266N1 & 713466N1 & 713666N1 & & \\[2mm]
0.76\Msolar & & & & & \\
         &          & 063476N1 & 063676N1 & 063876N1 & \\
         &          & 013476N1 & 013676N1 & 013876N1 & \\ 
         & 963676N1 & 963476N1 & 963676N1 & 963676N1  & \\
913076N1 & 913476N1 & 913476N1 & 913676N1 & & \\
863076N1 & 863276N1 & 863476N1 & & & \\
813076N1 & 813276N1 & 813476N1 & & & \\
763076N1 & 763276N1 & 763476N1 & & & \\
\hline
\end{tabular}

\end{table}

\subsection{Opacity}
\label{s:opacity}

Pulsation models are critically dependent on the opacity and hence on the assumed chemical composition in the stellar envelope (\S\,\ref{s:comp}). 
Opacity tables have been prepared using data from the OPAL calculations \citep{opal92,iglesias96}.
Since each opacity table covers a wide range of hydrogen and helium abundances, only the detailed
distribution of elements with $Z>2$ is necessary to build the tables. 

The primary composition adopted is based on that measured by \citet{jeffery01b} {\bf J01 $\equiv$ N1}.   
Here, the abundances of several elements including sodium, chlorine, argon, potassium, calcium, titanium, chromium, manganese and nickel are based on a scaled solar abundance relative to iron \citep{asplund09}. 
The normalized abundances by mass of elements with $Z>2$ are shown in Table~\ref{t:opacity}.

To investigate possible errors in the envelope metallicity (factors of two are within the experimental uncertainties), we also
prepared  tables  for mixtures with iron and nickel abundances enhanced by factors of two and three. 
These mixtures are labelled {\bf FeNi2 $\equiv$ N3} and {\bf FeNi3 $\equiv$ N6}. For mixture {\bf N6}, abundances of elements sodium, chlorine and calcium to manganese were also increased by $\approx50-80\%$. 

The same mixtures were also assumed for the composition of
the envelope, so that the equation  of state and opacity are as far as possible consistent. 

For all of our calculations, we have defined the hydrogen abundance (mass fraction) to be $X=0.00125$  \citep{przybilla05}. 
For the standard mixture \textbf{J01} the metal mass fraction $Z=0.0159$ \citep{jeffery01b}, and the helium mass fraction $Y=1-X-Z=0.98285$. 
For the iron and nickel enhanced mixtures \textbf{FeNi2} and \textbf{FeNi3}, $Z$ is increased to $0.0161$ and $0.01638$ respectively and  $Y$ is adjusted accordingly. 

A small number of calculations were carried out using OP opacities \citep{Seatonetal1994} instead of OPAL opacities. These assumed the \textbf{J01} mixture and were labelled \textbf{N5}.

%%====================================================================

\begin{figure}
\centering
\includegraphics[width=\columnwidth]{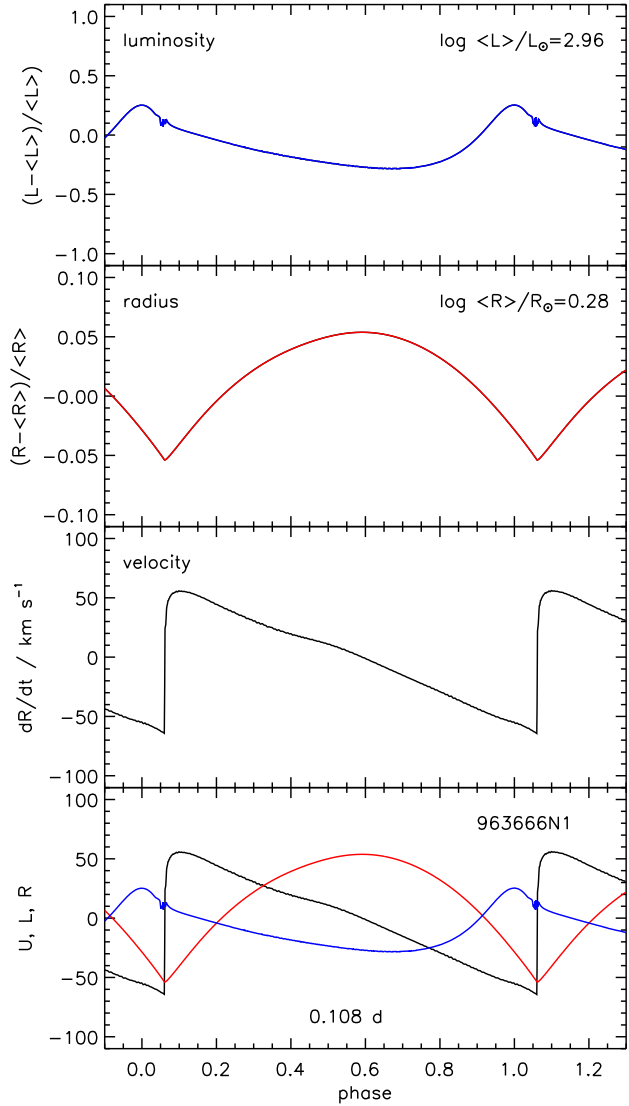}
\vspace*{-24mm}
\caption{The variation in luminosity ($L$, blue), radius ($R$, red) and radial velocity ($U \equiv {\rm d}R/{\rm d} t$, black) as a function of pulsation phase for a model with $\log L/\Lsolar=2.96$, $\log \Teff/{\rm K} = 4.36$, $M/\Msolar=0.66$ and mix N1. 
Phase zero is defined to be at maximum light. The period is shown in the middle of the bottom panel. 
The top three panels show each variable separately, with luminosity and radius shown as the fractional change relative to the mean, as indicated in the appropriate panel.
The bottom panel shows all three variables superimposed, with luminosity and radius rescaled by factors of 100 and 1000 respectively so that they can be more easily plotted a common scale. 
To be clear, a fractional change $\delta L/\langle L\rangle \equiv (L-\langle L\rangle)/\langle L\rangle) = 0.10 \equiv 10\%$ corresponds to a displacement 
of $+10$ on this scale.
Likewise, a fractional change $\delta R/\langle R\rangle \equiv (R-\langle R\rangle)/\langle R\rangle) = 0.01 \equiv 1\%$ also corresponds to a displacement of $+10$.
Such panels are used at reduced sizes in subsequent figures to illustrate the variation of pulsation behaviour with $L$, $\Teff$ and $M$.
}
\label{f:96636N1}
\end{figure}

%\begin{landscape}
\begin{figure*}
\centering
\vspace*{-15mm}
\includegraphics[width=1.0\textheight,angle=90]{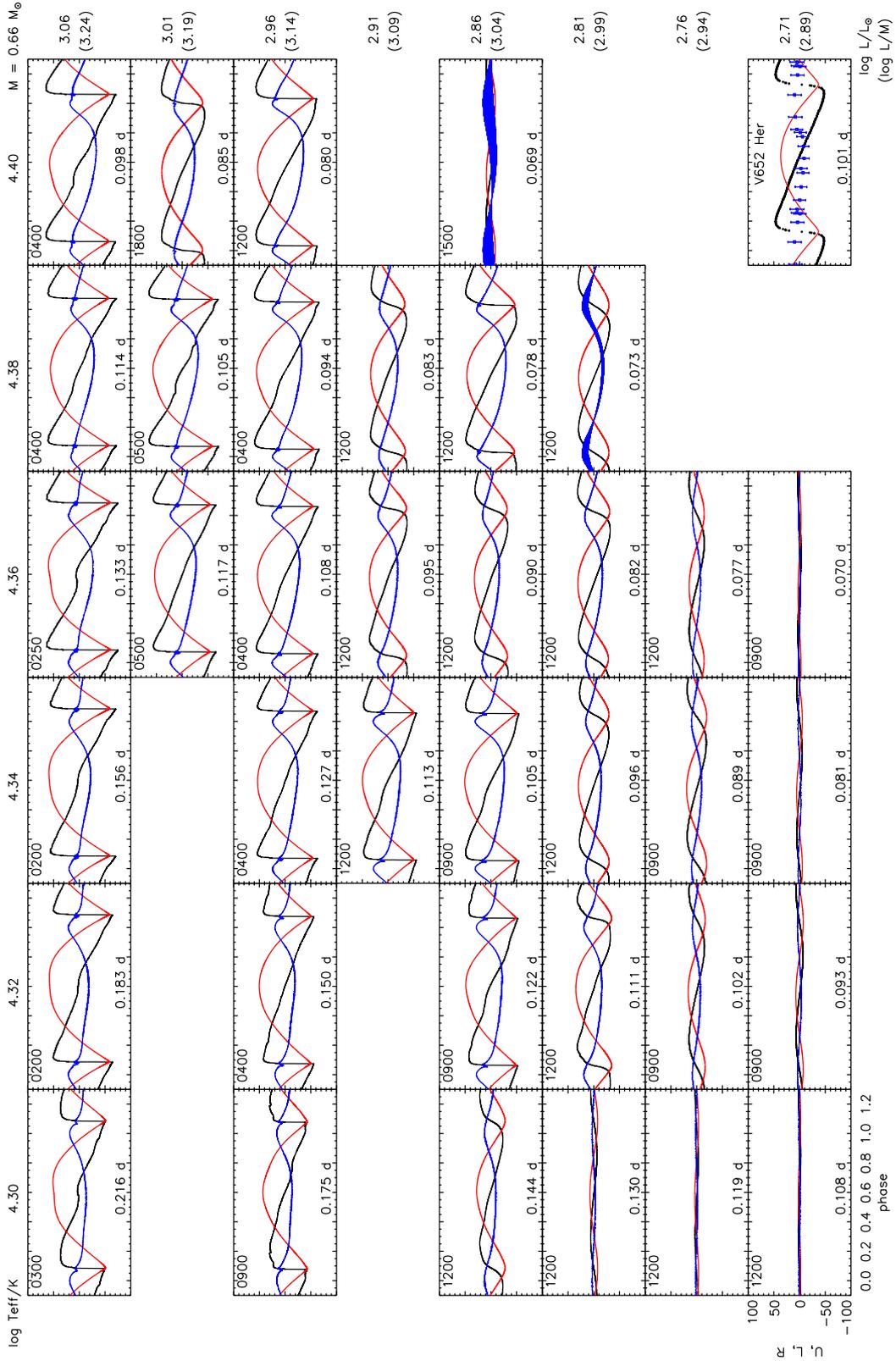}
\vspace*{-12mm}
\caption{Luminosity ($100 \delta L/\langle{L}\rangle$, blue), radius ($1000 \delta R/\langle{R}\rangle$: red) and radial velocity ($U \equiv {\rm d}R/{\rm d} t / \kmsec$: black) as a function of phase for models with $M/\Msolar=0.66$ and mix N1. The input parameters $M$, \Teff,  $L$ (and $L/M$) associated with each model are shown along the top and right hand axes respectively. Phase zero is defined to be at maximum light. The period for each model is shown at the bottom of each panel. The number shown top left of each panel represents the number of runs completed for each model, each run representing 24\,000 time steps. Similar panels for models with  $M/\Msolar=0.56$ and 0.76 and over an extended range of \Teff and  $L$ are given in the Appendix. 
The axes and scales for all panels are identical; for clarity only axes in the lower-left panel are labelled.
The bottom right panel shows the equivalent quantities as observed in V652\,Her and reported in Figs.\,3 and 7 of \citet{jeffery15b}, assuming a mean radius $\langle{R}\rangle=2.31\Rsolar$ (Table\,\ref{t:pars}). 
}
\label{f:mods66}
\end{figure*}
%\end{landscape}

%--------------------------------------------------------------
\section{periods, surface and internal properties}
\label{s:periods}

 Figure \ref{f:96636N1} shows the behaviour of radius, luminosity and radial velocity as a function of phase for the representative model 963666N1. 
It explains the scales and demonstrates how the individual panels in 
Fig. \ref{f:mods66} and Figs.\,\ref{f:mods56}-\ref{f:66N3} are constructed. 
The latter show the radii, luminosities and radial velocities as a function of pulsation phase for many models of a given mass and chemical mixture. 
Within each figure, individual panels show results on the grid of \Teff\ and $L$ defined above. 
To limit computational time, work focused on models having pulsation period close to 0.11\,d, with some extension to higher and lower \Teff\ to explore systematics. 
Phase has been defined to be zero at luminosity maximum on the assumption that this is co-eval with maximum light in the Johnson V-band \citep{landolt75}. This assumption should be verified.  

Figure \ref{f:mods66} and Figs.\,\ref{f:mods56}-\ref{f:66N3} demonstrate how shape, amplitude, and phase-shift vary as functions of model parameters. Data used to prepare these figures, sampled at 600 phase points, together with procedures for reading and displaying them, are available as supplementary online material (Table\,\ref{t:supp}); their format is illustrated in Table\,\ref{t:surface}. 
Most models show the characteristic saw-tooth radial velocity curve observed in V652\,Her (Fig.\,\ref{f:mods66}: lower right), with a steep acceleration phase lasting $\sim0.1$ cycles and a slow and nearly-constant deceleration phase lasting $\sim0.8$ cycles.
There are three principal features.
1) For models having the same mass and luminosity, the amplitude of the oscillation increases with effective temperature, with some levelling off at  $\log \Teff/{\rm K}\gtrsim 4.35$. 
There is some evidence that the amplitude starts to drop for $\log \Teff/{\rm K}\gtrsim 4.40$, but model convergence becomes increasingly difficult beyond this value.
2) For models having approximately the same period, the velocity amplitude increases with both effective temperature and luminosity, since the two are tied. The same limiting amplitude applies.   
3) Structure appears in the deceleration phase of some radial-velocity curves. 
This appears to be related to amplitude and hence to opacity, since these secondary structures are more common in a grid computed with iron and nickel enhanced chemistry (mixture {\bf N3}, Fig.\,\ref{f:66N3}).

Figures \ref{f:guts} and \ref{f:surf} represent the internal displacement and temperature of layers close to the surface through the pulsation cycle for models 963666N1. 
The temperature is represented by a colour scale.

%%====================================================================

\begin{figure*}
\centering
\includegraphics[width=0.40\linewidth]{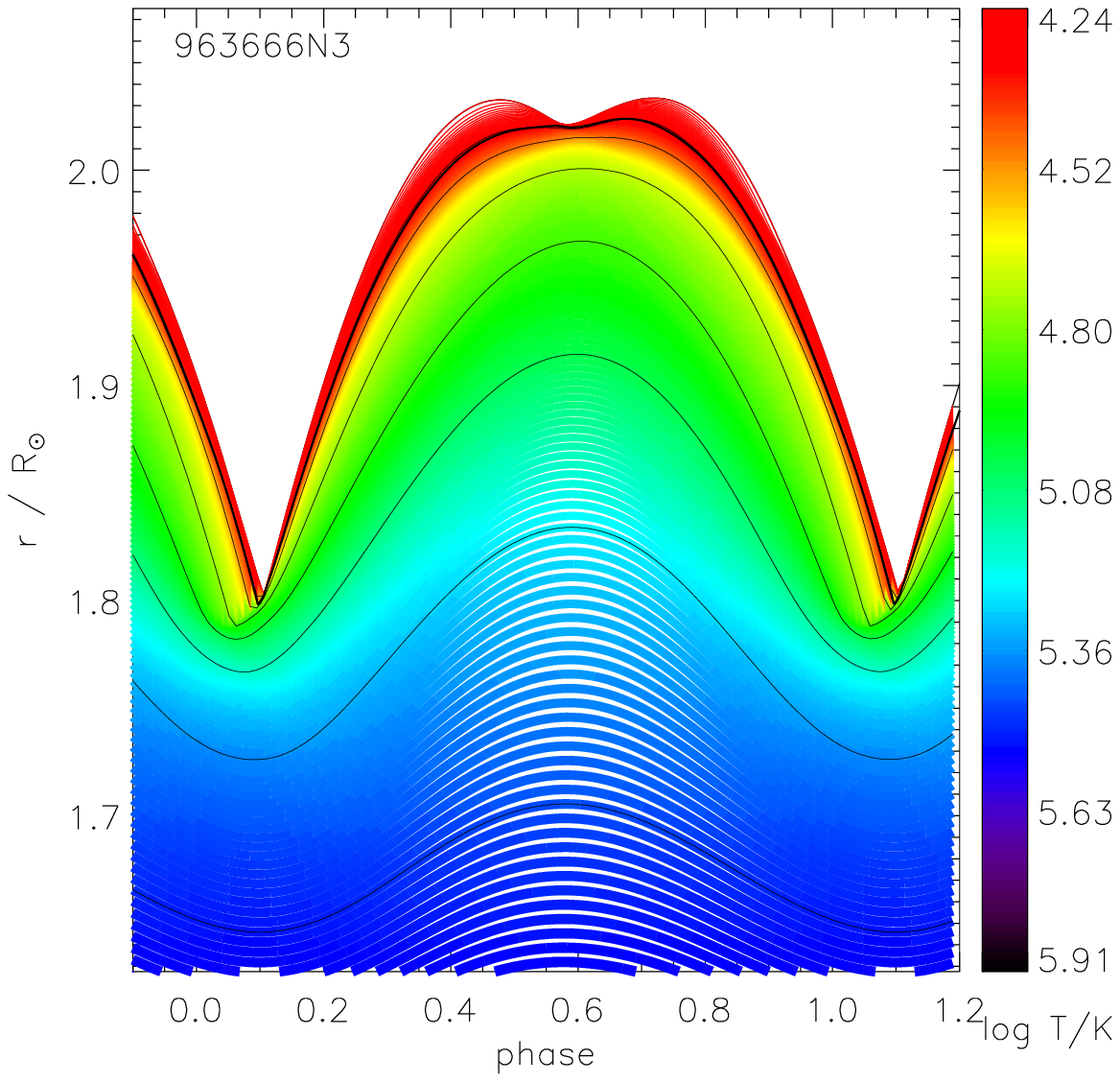}
\includegraphics[width=0.40\linewidth]{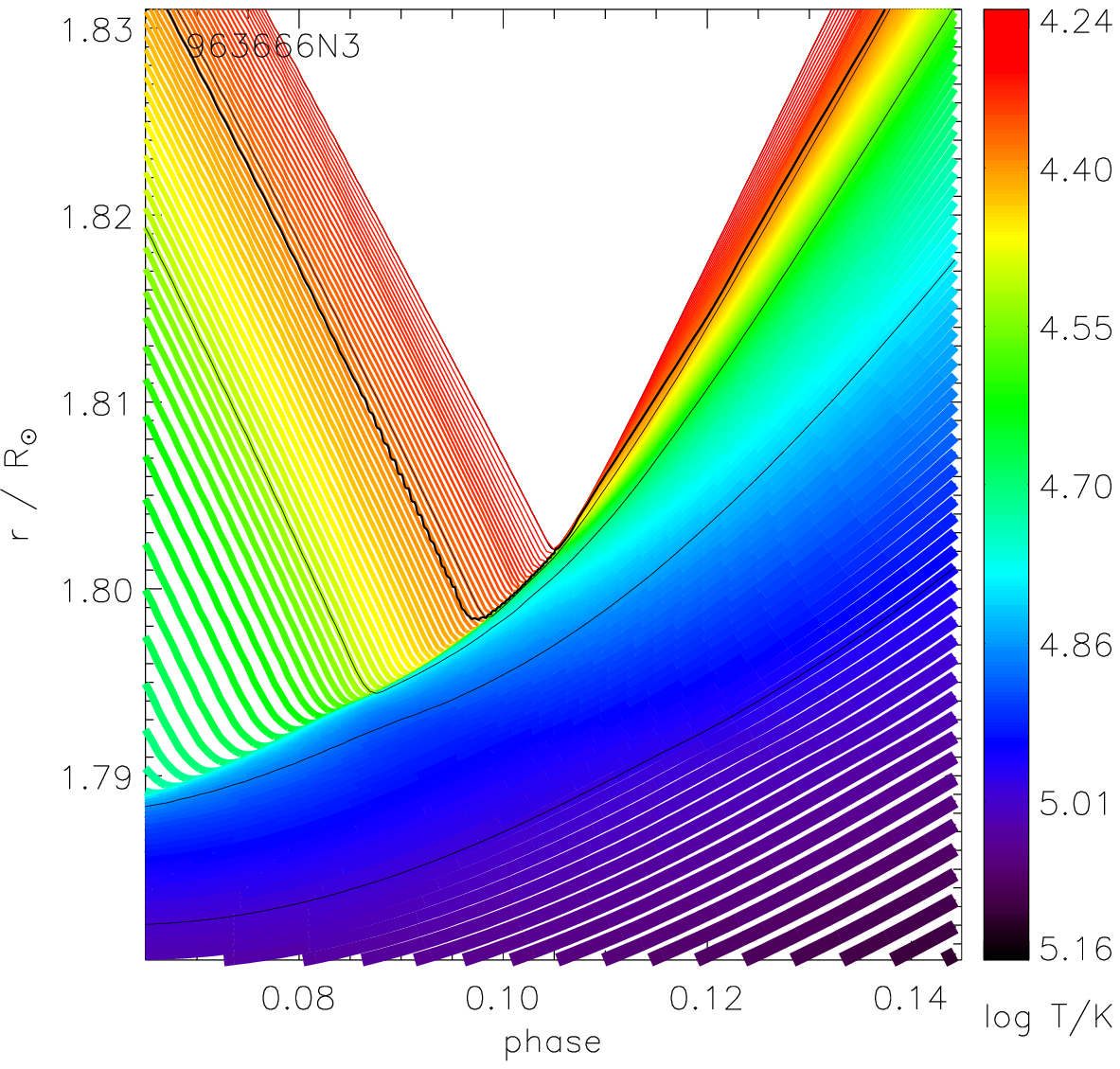}
\\
\includegraphics[width=0.40\linewidth]{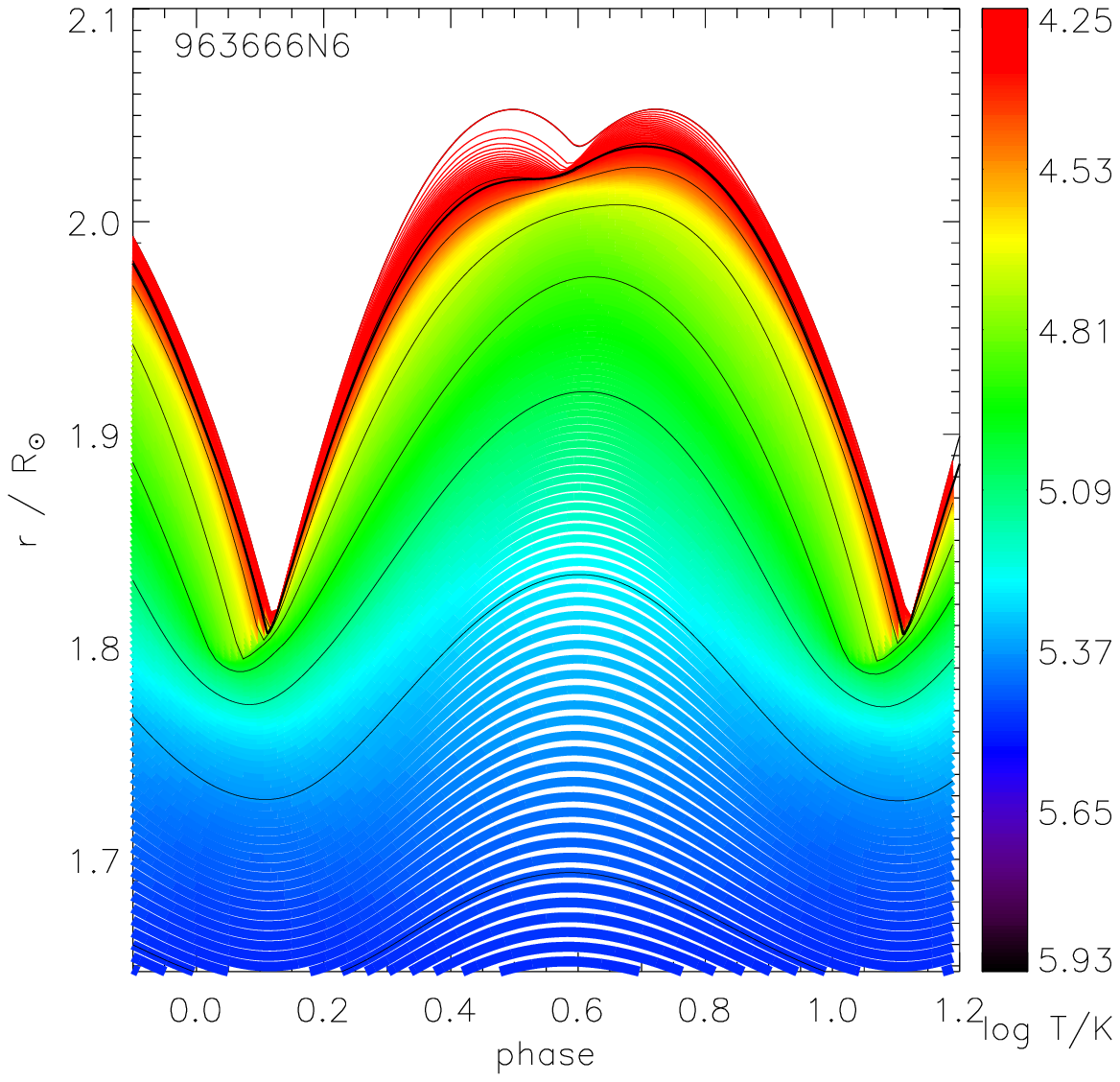}
\includegraphics[width=0.40\linewidth]{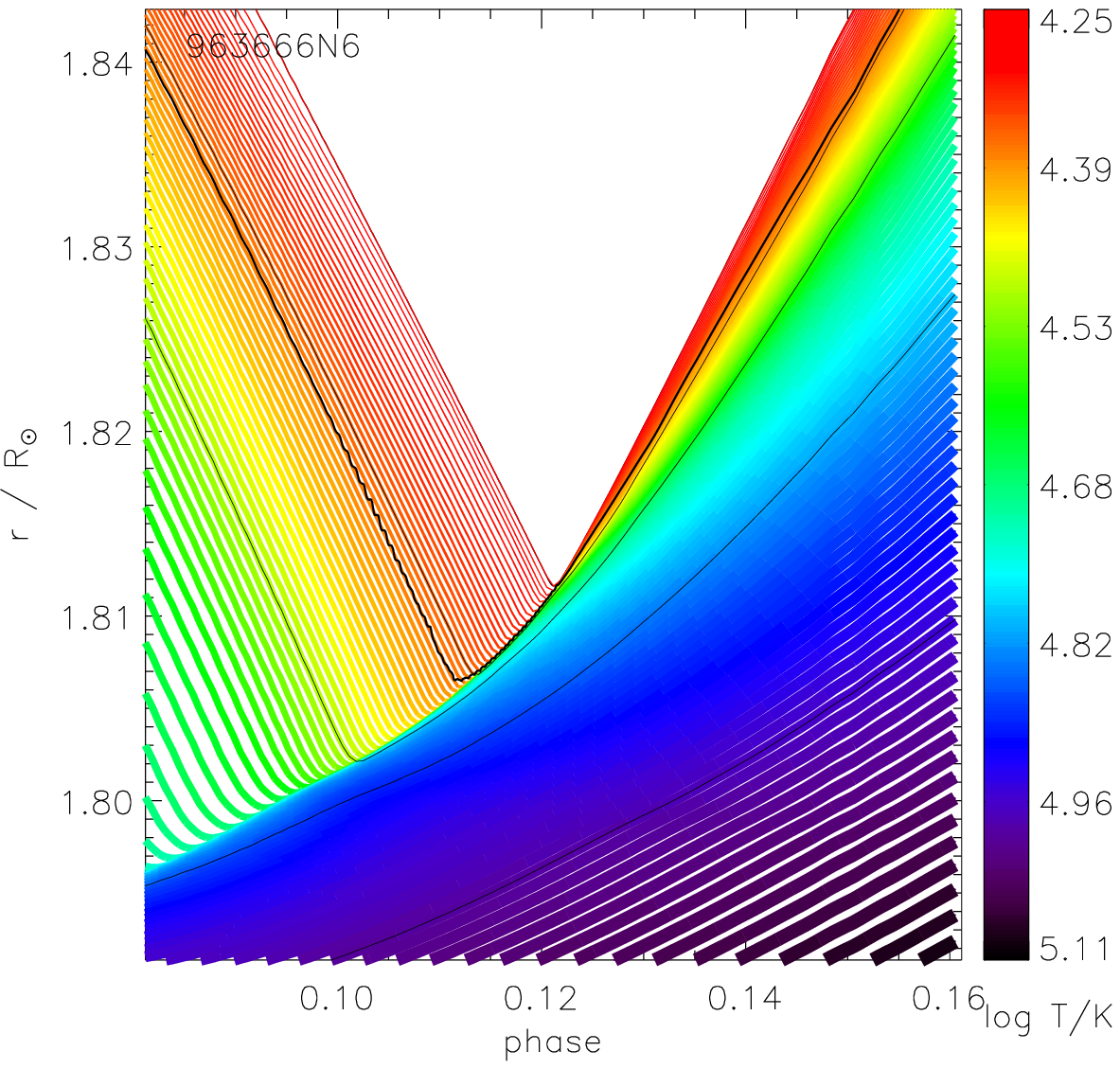}
\\
\includegraphics[width=0.40\linewidth]{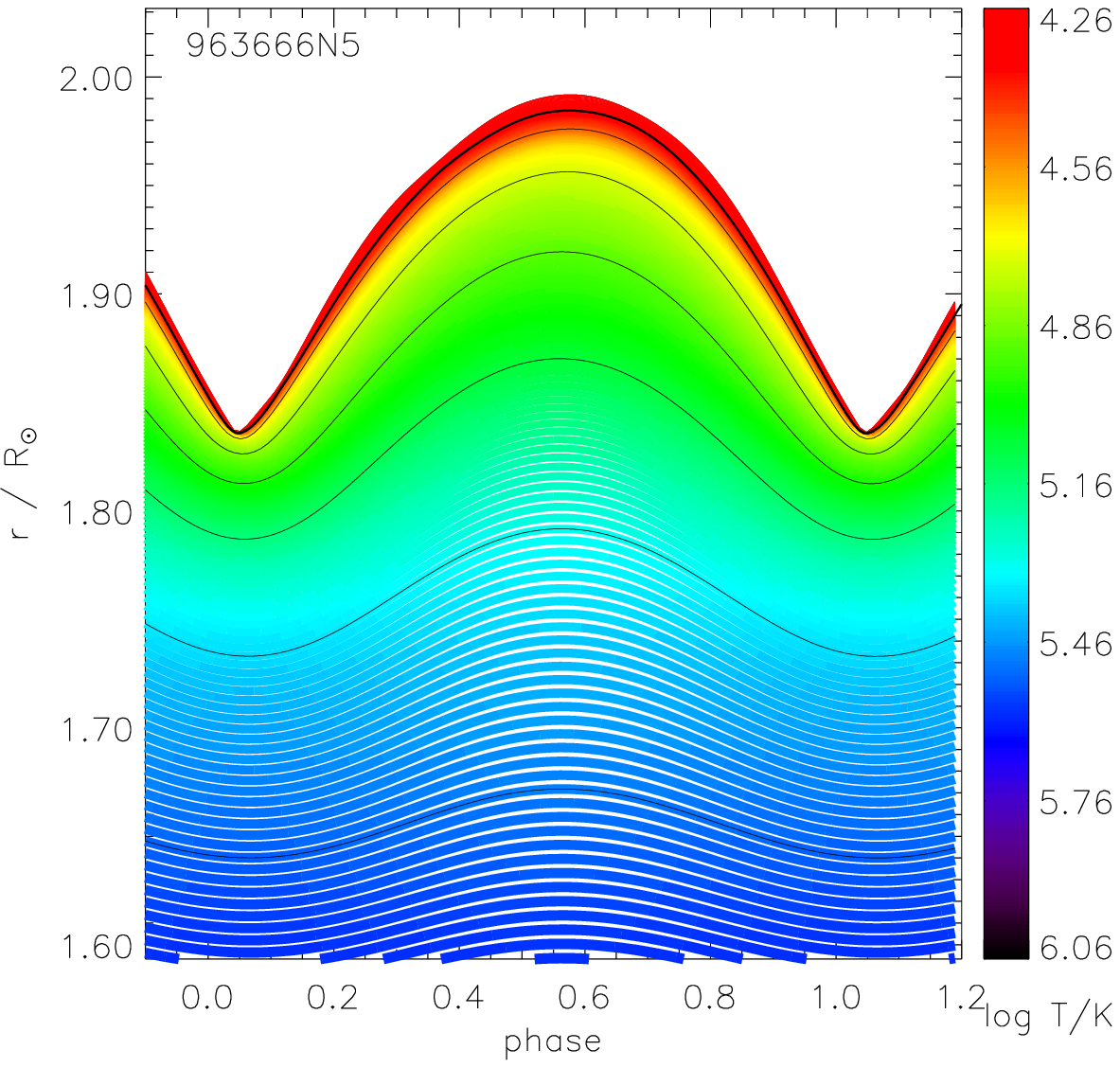}
\includegraphics[width=0.40\linewidth]{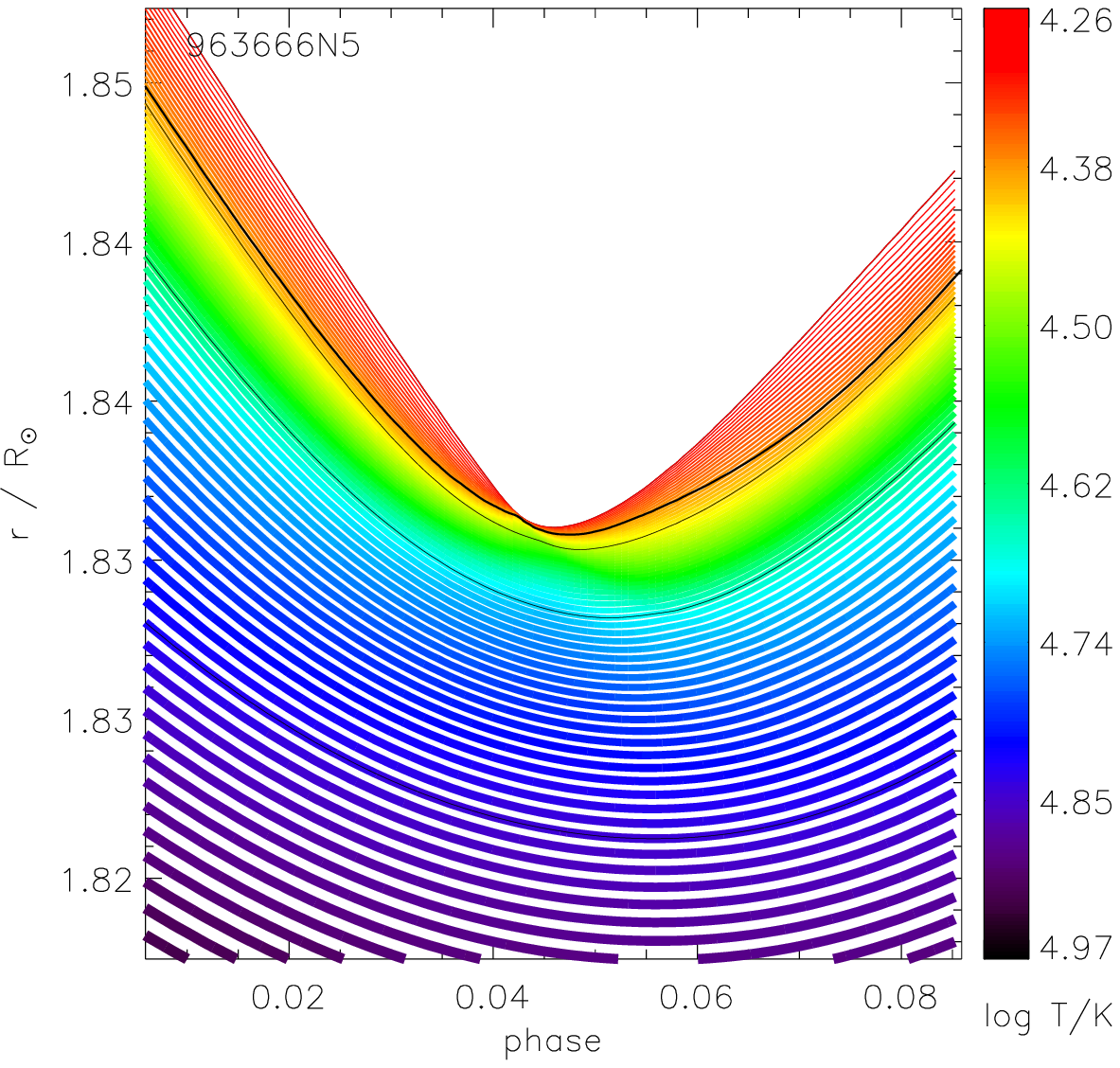}
 \caption{As Fig.\ref{f:surf} for mixtures with $2\times$ (N3: top) and $3\times$ (N6: middle) the iron and nickel abundances and with OP opacities (N5: bottom). 
} 
	\label{f:opguts}
\end{figure*}

\begin{figure*}
\centering
\includegraphics[width=0.40\linewidth]{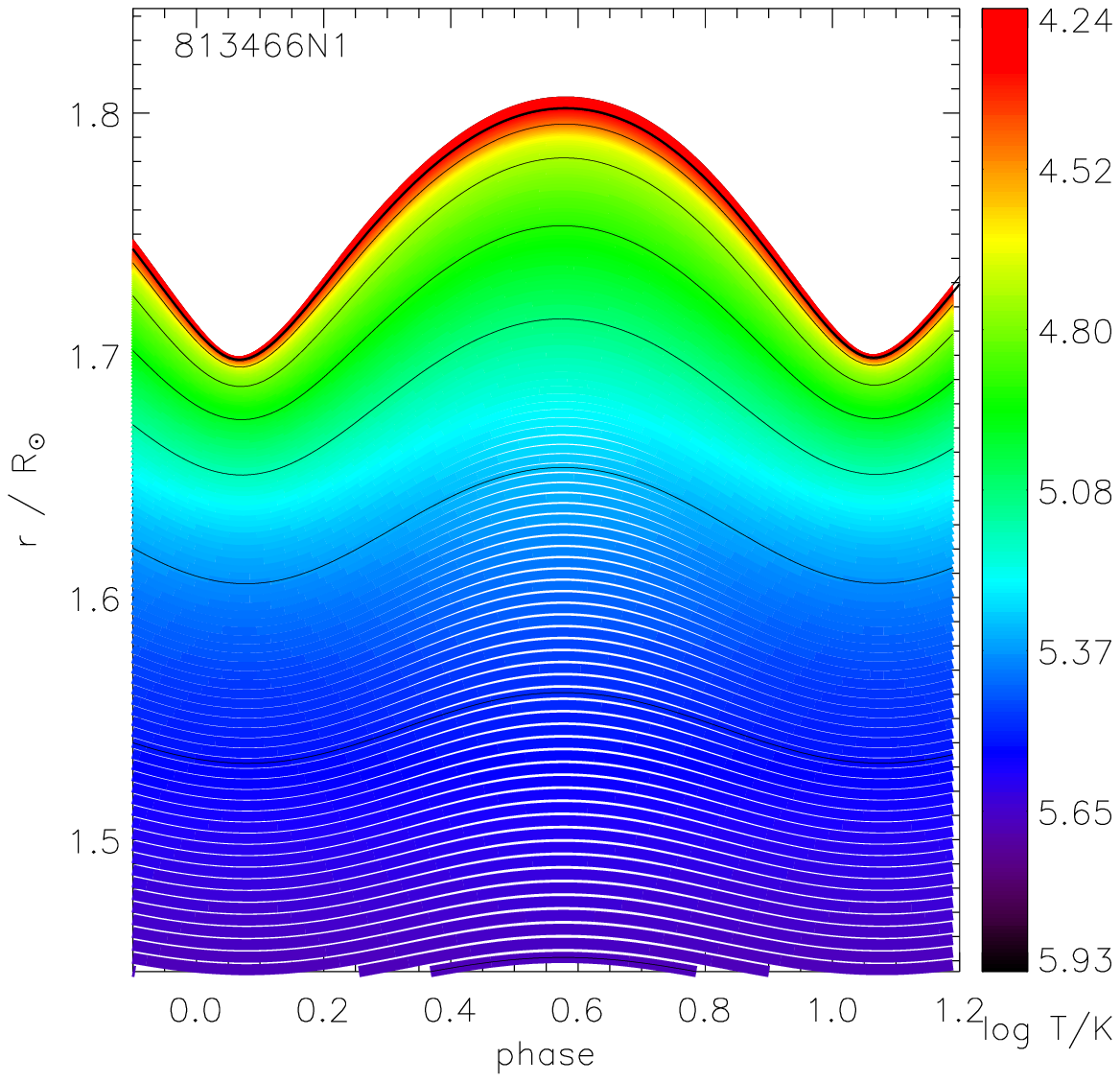}
\includegraphics[width=0.40\linewidth]{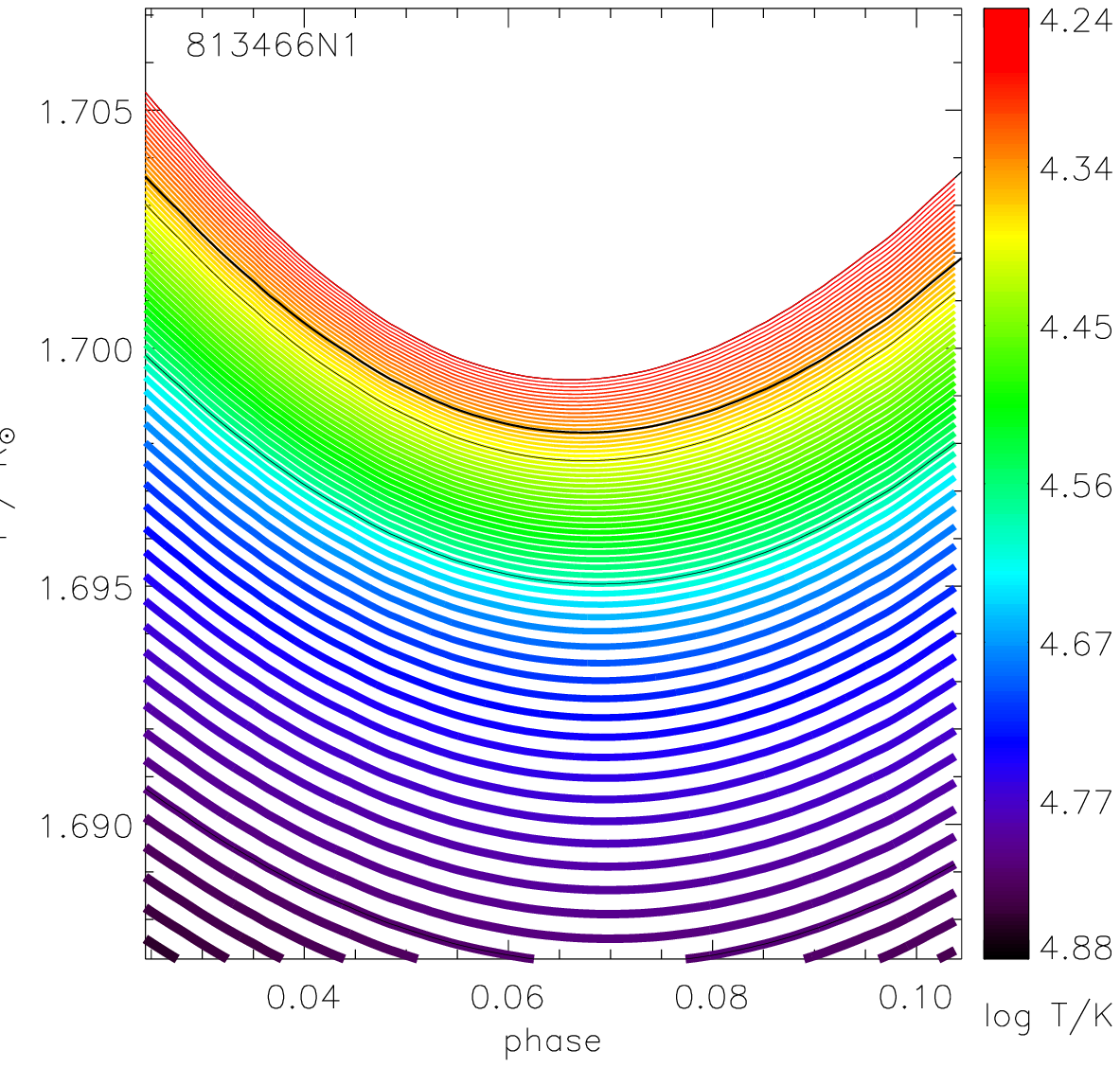}
\\
\includegraphics[width=0.40\linewidth]{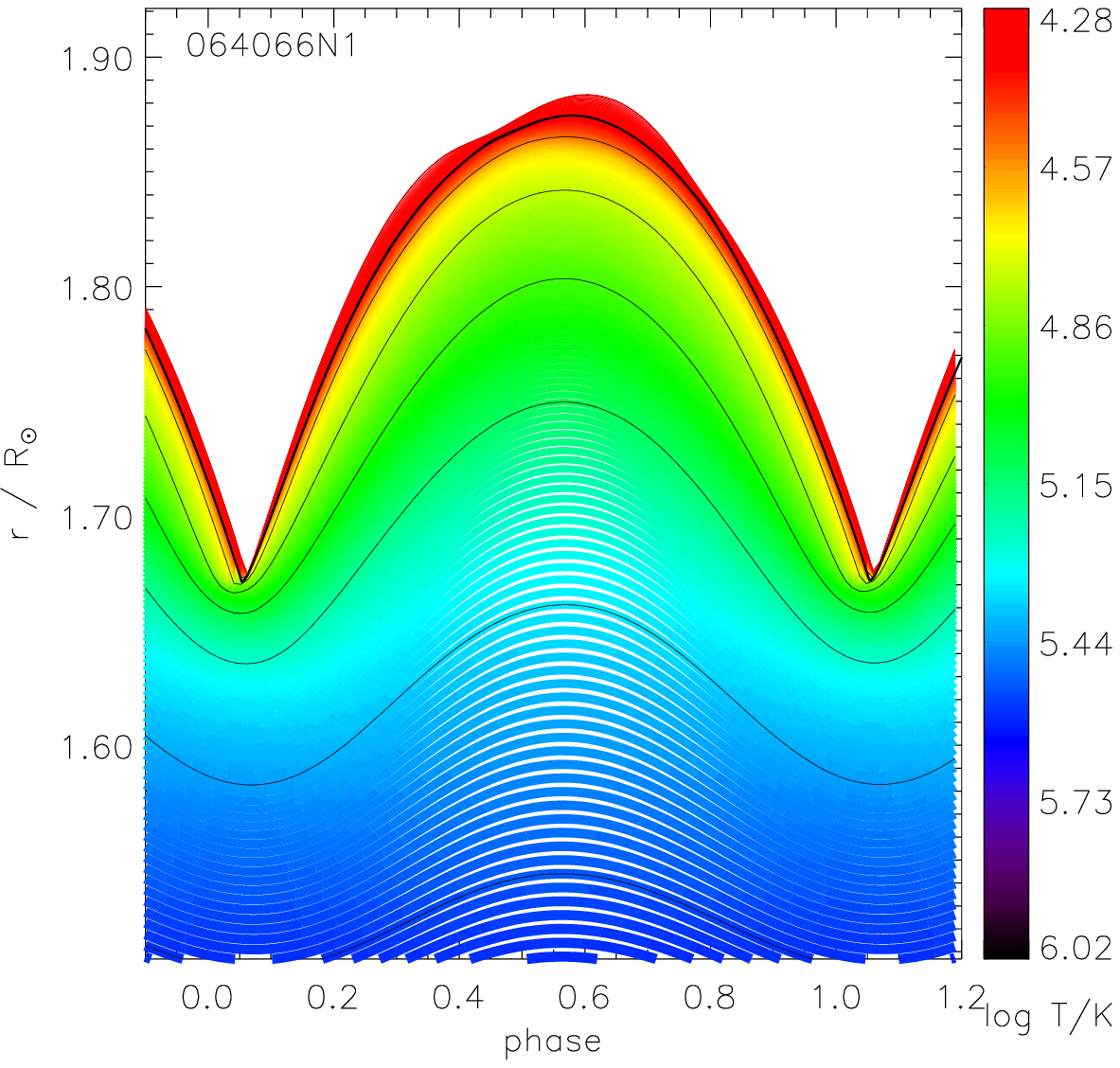}
\includegraphics[width=0.40\linewidth]{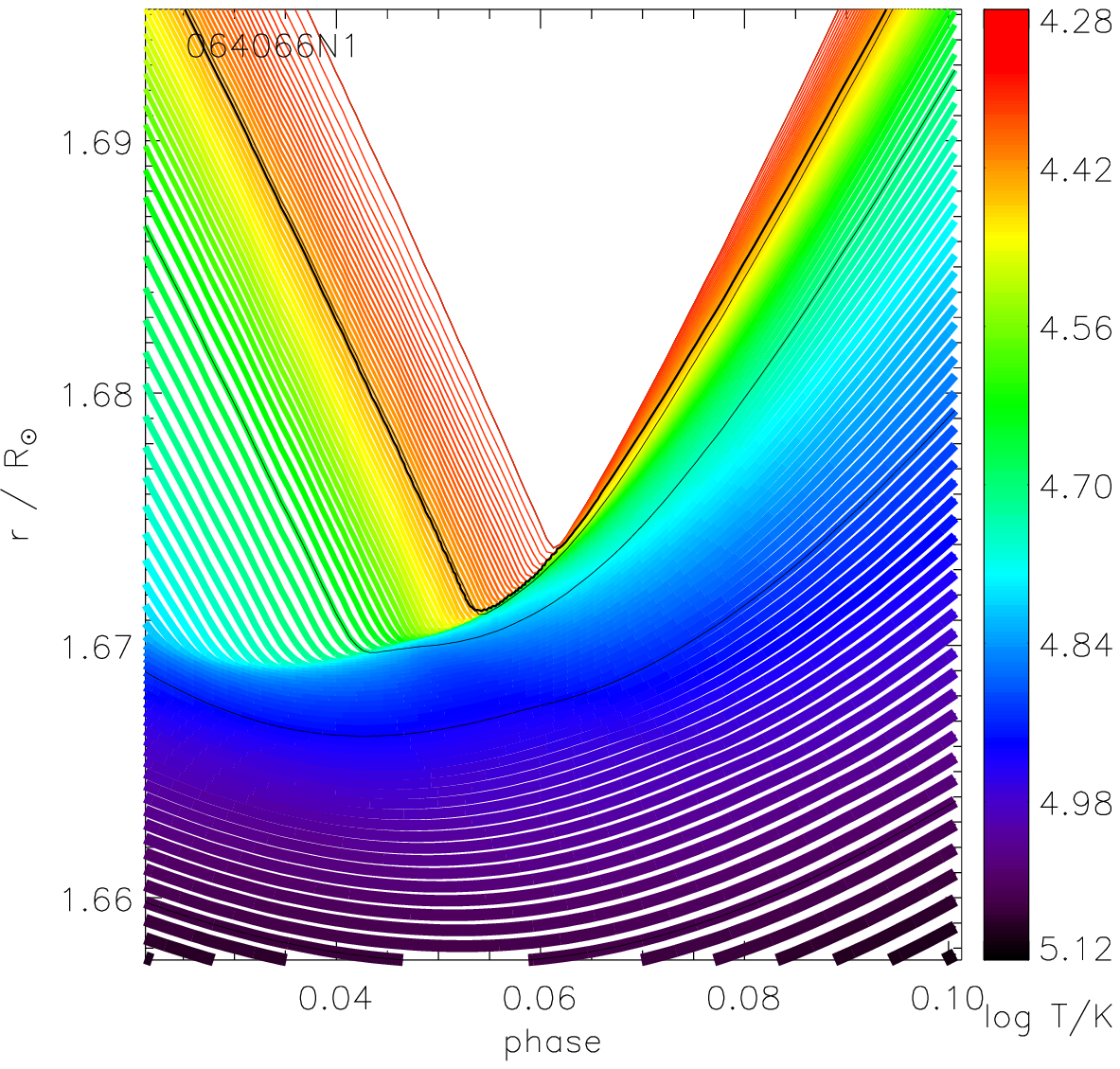}
 \caption{As Fig.\ref{f:surf} for $0.66 \Msolar$ models with with $\log L/\Lsolar=2.81$ (upper panels) and 3.06 (lower panels). 
} 
	\label{f:opguts2}
\end{figure*}

\subsection{Opacity effects}

Using model 963666N1 as a reference, calculations were carried out to test the effect of varying the opacity. Mixtures N3 and N6 are equivalent to mixture N1 except that the iron and nickel abundances are increased by factors of 2 and 3 respectively. Mixture N5 is the same as mixture N1 except that OP opacities were used in place of OPAL opacities. 

Figure \ref{f:opguts} compares the internal structure for the three test cases with the reference model (Fig.\,\ref{f:surf}).  
As anticipated, the increased iron and nickel abundances produce higher opacity, a steeper opacity gradient and hence stronger driving. 
The primary consequences are a larger amplitude pulsation, faster acceleration at minimum radius, and a profoundly stronger shock which commences earlier and deeper within the envelope. 

In this comparison, we also find that the phase delay between maximum luminosity and minimum radius increases with increasing iron and nickel abundance and hence with iron-bump opacity. 
As a consequence we recomputed the 0.66\,\Msolar\ grid using mixture {\bf N3} (iron and nickel enhanced $\times 2$) (Fig.\,\ref{f:66N3}). 

A third consequence is, at least in our models, a separation of the atmosphere at minimum radius into multiple components, the outer being accelerated more rapidly than the inner.
The additional acceleration is a consequence of shock heating, followed by adiabatic cooling.  
Thereafter, the outer layers decelerate more rapidly and are re-accelerated by the still expanding inner layers, so that the radius and radial-velocity curves are double humped.   
However, a limitation of the calculation is that strong shocks are not treated accurately; since the Courant condition is violated it is not clear that such behaviour would occur in real stars.
Hence the near-constant deceleration observed in V652\,Her might not provide an upper constraint on the iron+nickel abundance. 

The slightly larger amplitude obtained from the model using OP opacities implies the latter to be slightly larger than the OPAL opacities as implemented.  

\subsection{Luminosity  effects}

Figure\,\ref{f:opguts2} compares models with the same composition and roughly similar pulsation period but with lower and higher luminosities than the reference model, i.e. with $\log L/\Lsolar = 2.81$ and 3.06 
(models 813466N1 and 064066N1, $\log L/M / (\Lsolar/\Msolar) = 2.99$ and 3.24). 
Including the reference model, the radial amplitude  ranges from 0.11\Rsolar\ (813466N1),  through 0.16  \Rsolar\  (963666N1)  to  0.21  \Rsolar\ (064066N1). 
Since the $L/M$ range covered by all three models is less than a factor 2 ($\delta \log L/M = 0.25$), pulsation amplitude is clearly strongly correlated with luminosity.

\begin{figure*}
\centering
\includegraphics[width=0.48\linewidth]{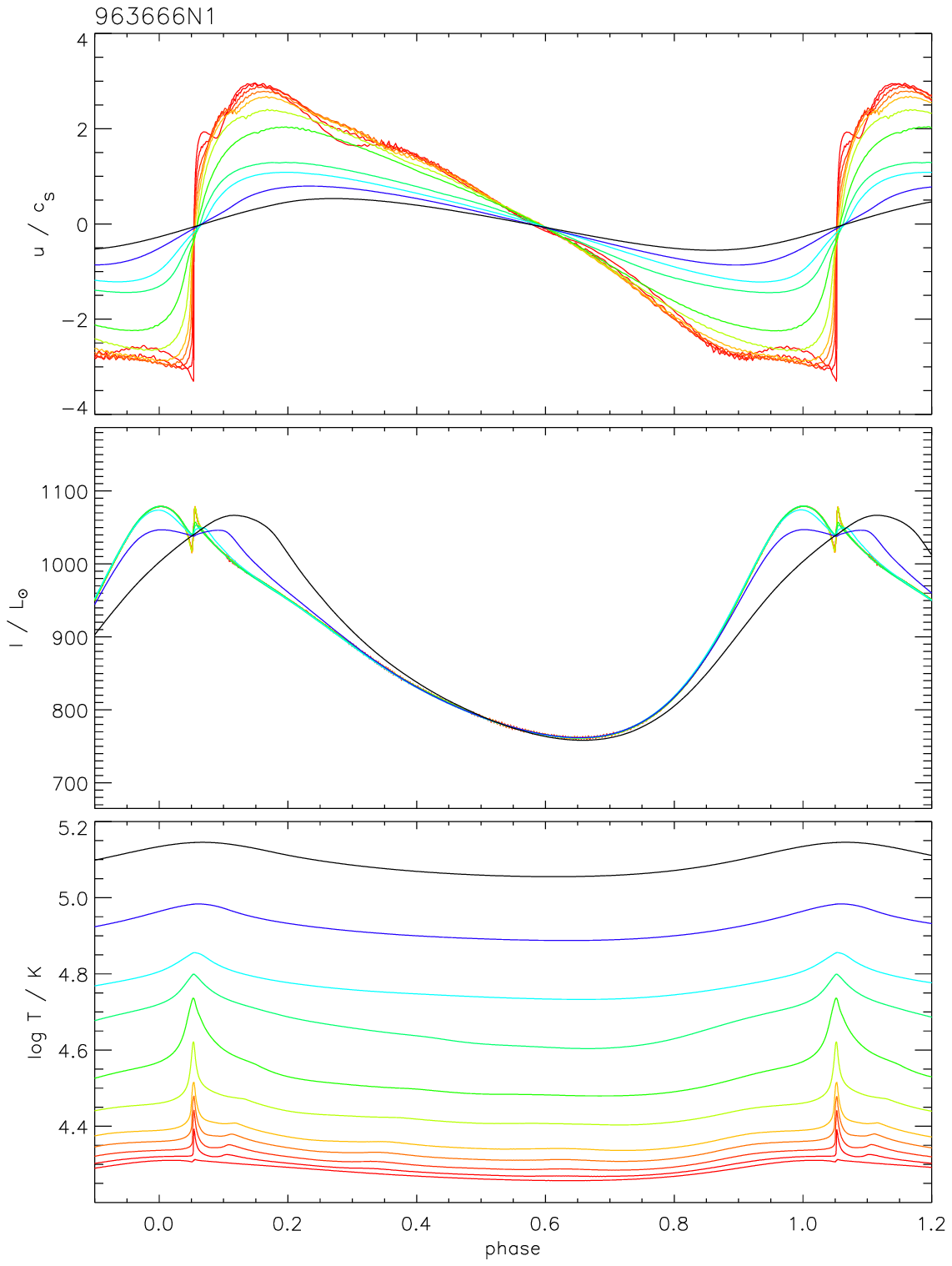}
\includegraphics[width=0.48\linewidth]{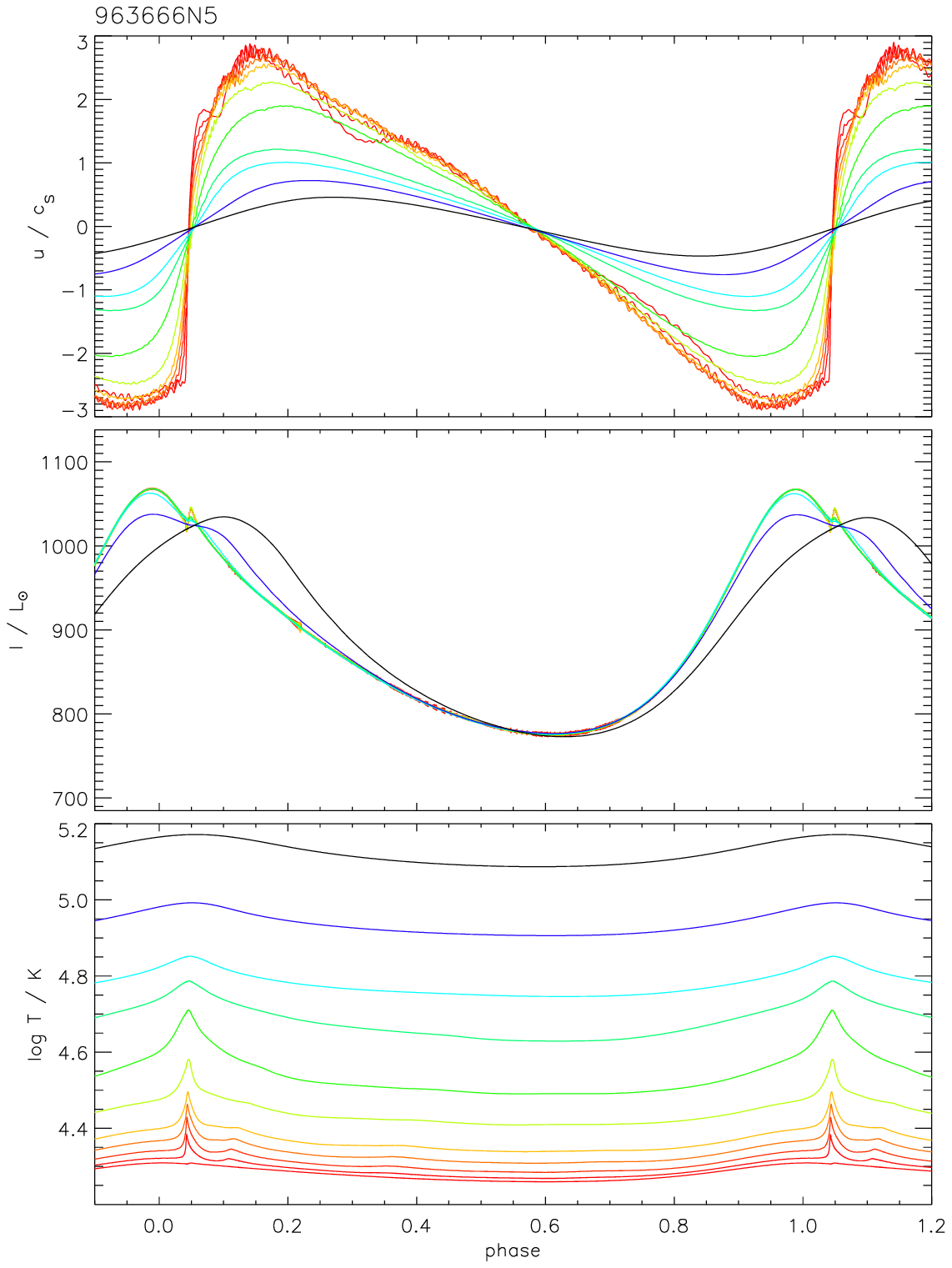}
\\
\includegraphics[width=0.48\linewidth]{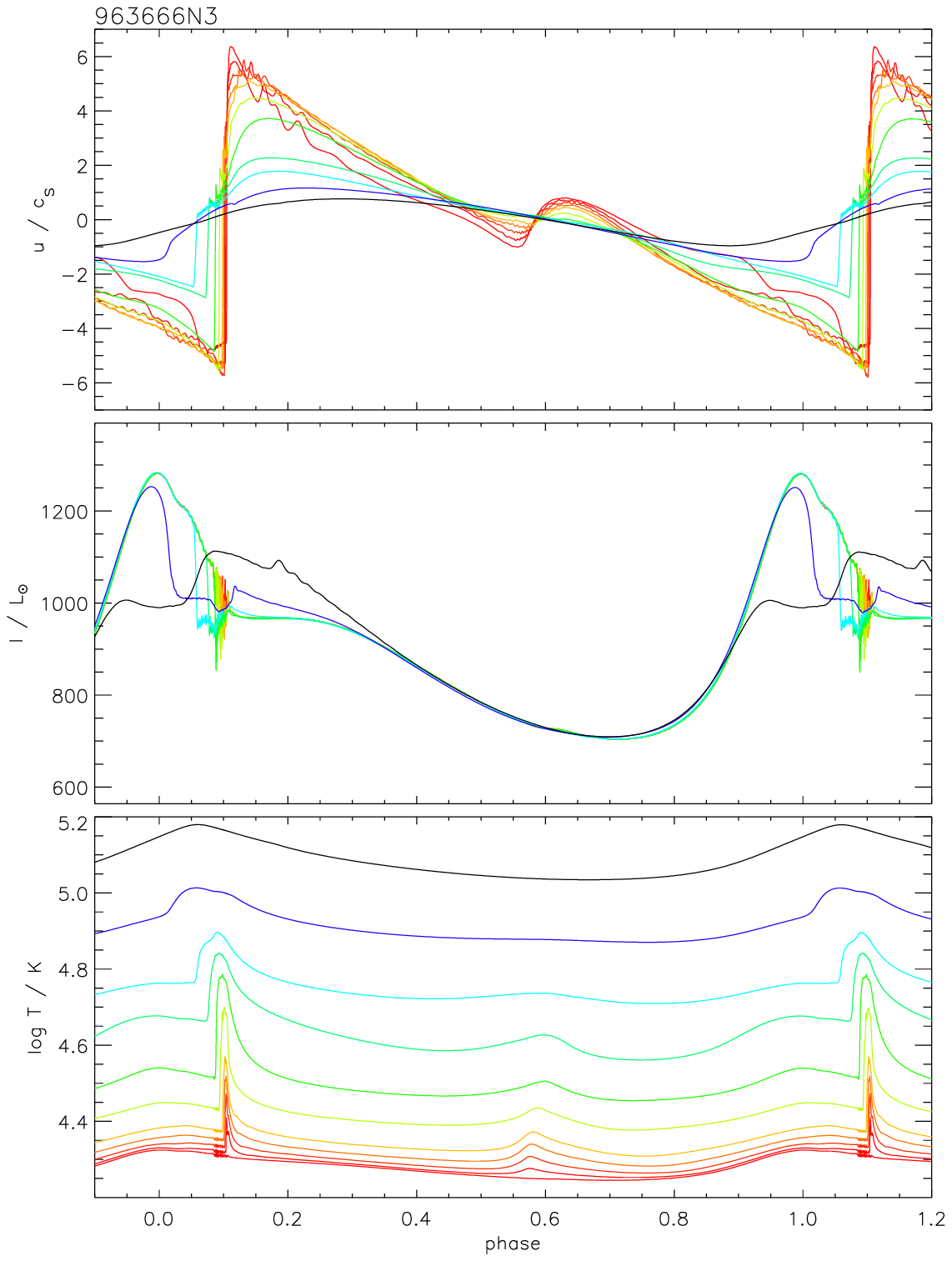}
\includegraphics[width=0.48\linewidth]{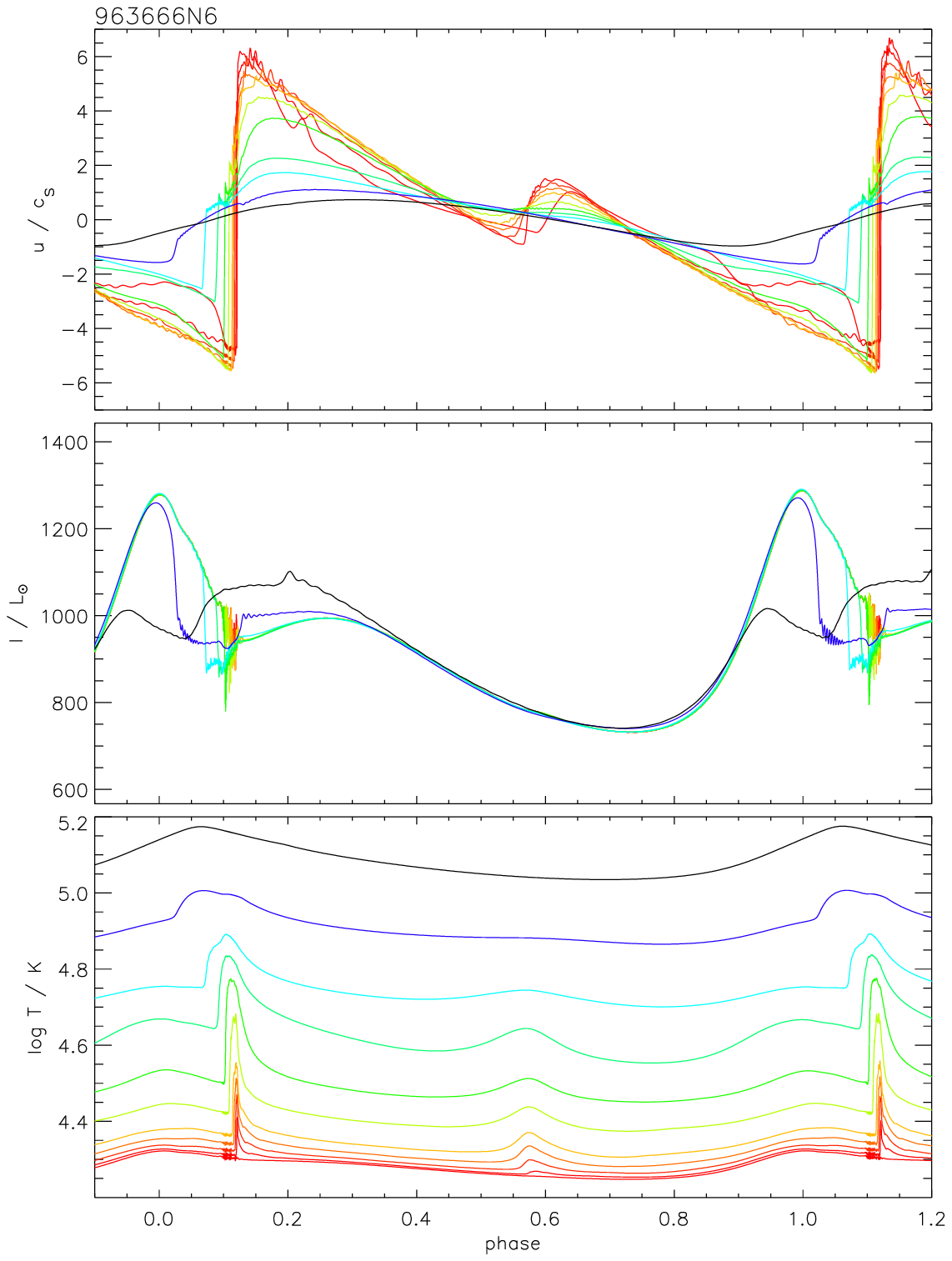}
\caption{
The runs of radial velocity $u$, luminosity $l$ and temperature $T$ with phase for selected zones in the outer part of the stellar envelope. Velocity has been scaled to the local sound speed. Each shell is colour-coded identically in each panel, red being coolest and violet (black) being warmest. The panels include the reference model (963666N1: top left), and with mixtures having $2\times$ (N3: bottom left) and $3\times$ (N6: bottom right) the iron and nickel abundances and with OP opacities (N5: top right). }
	\label{f:layers}
\end{figure*}

\begin{figure*}
\centering
\includegraphics[width=0.48\linewidth]{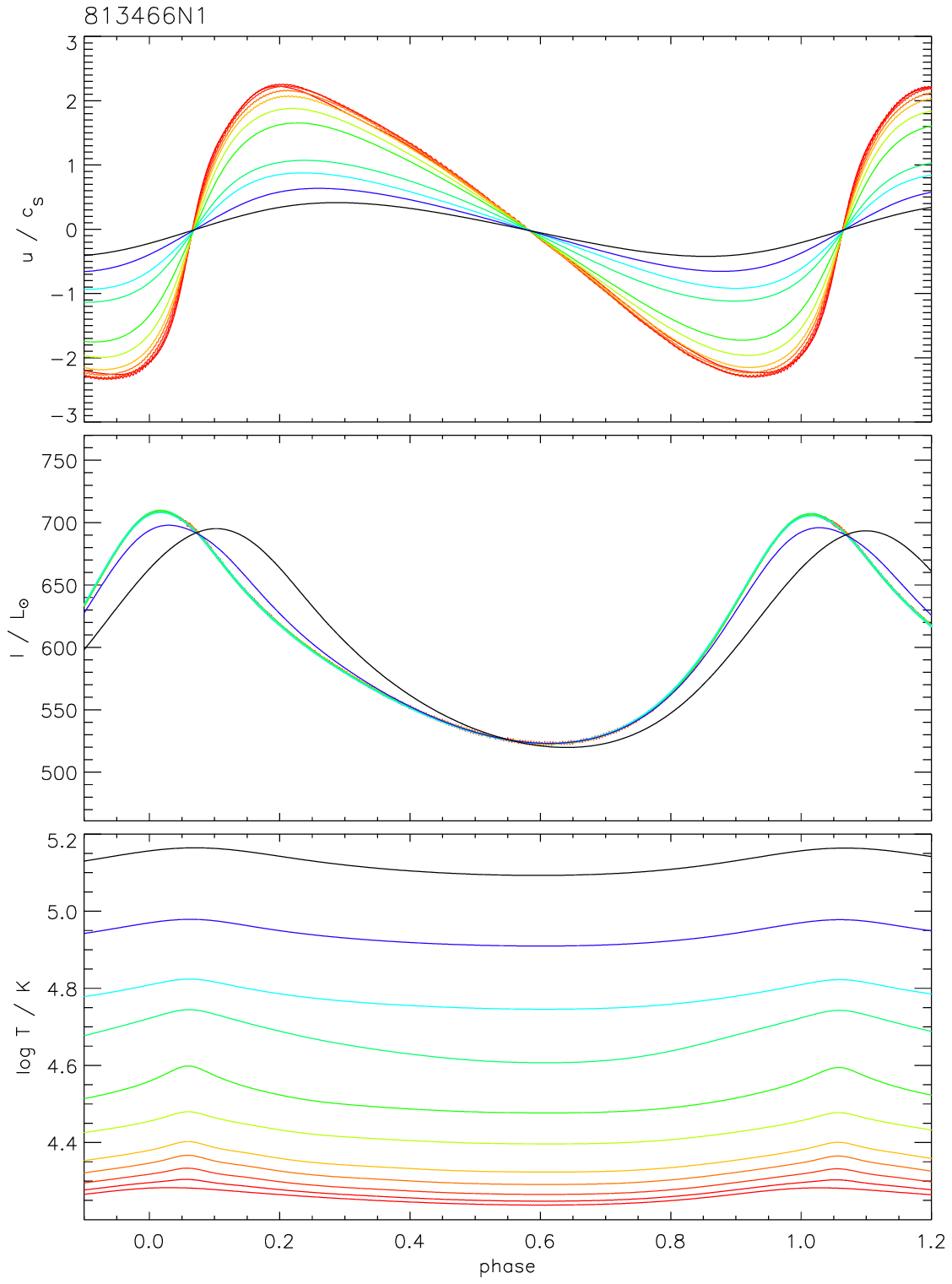}
\includegraphics[width=0.48\linewidth]{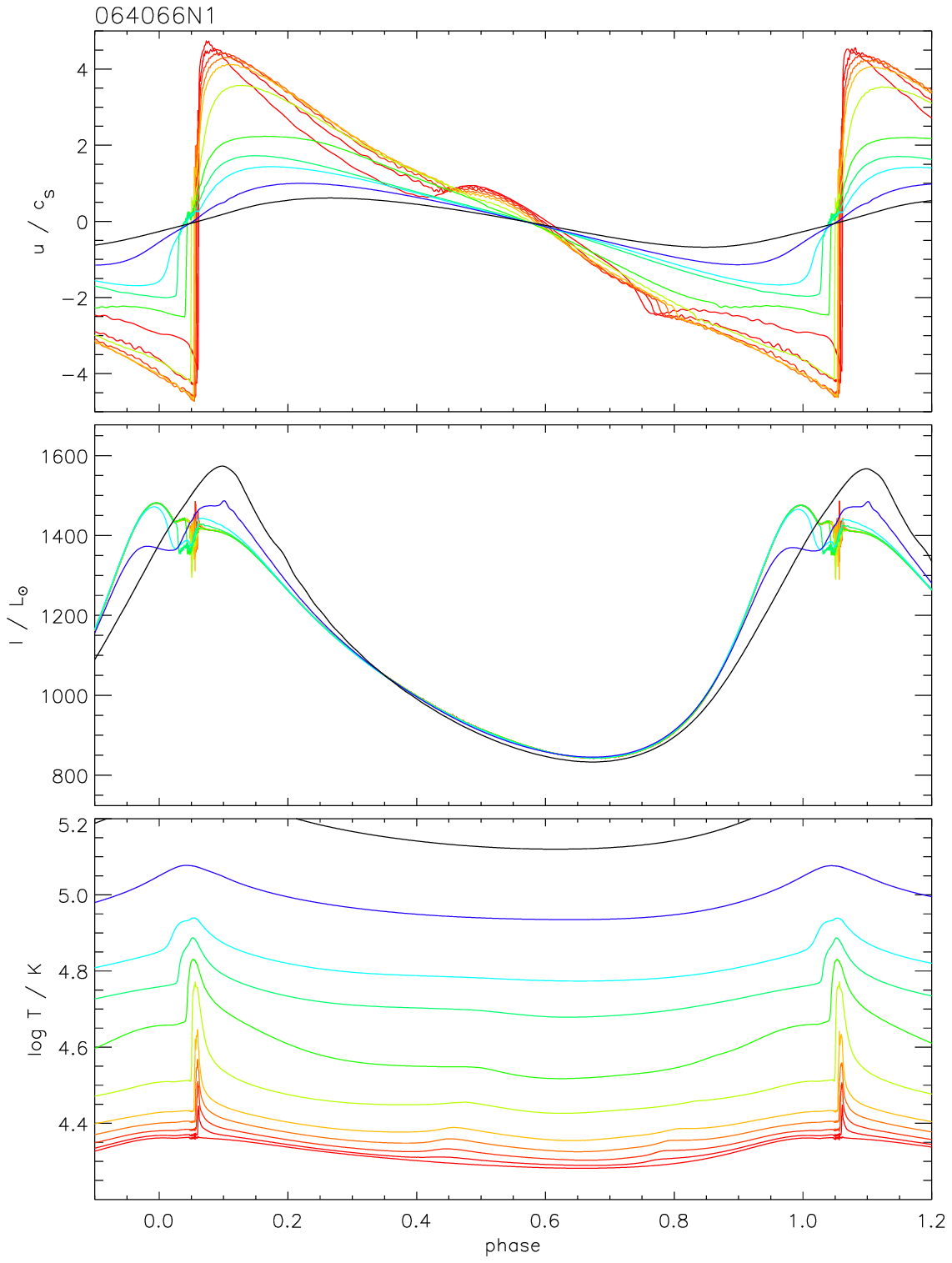}
\caption{
As Fig.\,\ref{f:layers} for the low and high $L/M$ models shown in Fig.\,\ref{f:opguts2}  }
	\label{f:layers2}
\end{figure*}

%%====================================================================

\subsection{Shocks} 

\citet{jeffery01b} argued that the surface of V652\,Her must be heated by the passage of the strong shock at minimum radius. 
Figures \ref{f:opguts} and \ref{f:opguts2} show a very steep change in temperature running outward through the star, commencing shortly before minimum radius, in models 963666N3, 963666N6 and 064066N1. 
A radial pulsation is a pressure wave propagating outward from the driving zone into a region of lower density (and sound speed).
Due to adiabatic compression and the consequent increase in sound speed, the compression front will catch up with the preceding pressure trough; if the former overruns the latter, a shock forms. 
So factors affecting shock formation include the amplitude of the pulsation and the sound-speed gradient in the stellar envelope.   
From Figs.\,\ref{f:guts} and \ref{f:opguts2}, 
no shock is generated in the low $L/M$ model (813466N1), the surface layers of the reference model (963666N1) are highly compressed at minimum radius (Fig.\,\ref{f:guts}), and the high $L/M$ model (064066N1) shows a strong shock. 

Figures \ref{f:layers} and \ref{f:layers2} present the models of Fig.\,\ref{f:surf},  \ref{f:opguts} and \ref{f:opguts2} in a different format. 
For each model, three panels show the radial velocity relative to the local sound speed\footnote{$c_s = \sqrt{\gamma P / \rho}$, assuming  $\gamma=5/3$.}, luminosity and temperature of selected zones.

Note  the behaviour of  model 813466N1 ($M=0.66\Msolar, \log \Teff/{\rm K}=4.34, \log L/\Lsolar = 2.81$)  in Figs.\,\ref{f:opguts2} and \ref{f:layers2}. 
In this relatively low $L/M$ model, surface motion is almost sinusoidal and there are no discontinuities in the radial velocity or temperature at any point in the models. 
Temperature maximum occurs almost simultaneously at all layers, except the outermost, where it coincides with light maximum and precedes radius minimum. 
 
For models with higher $L/M$, the development of a shock at minimum radius
is signalled by a spike in temperature  which corresponds to shock heating.
For milder cases (e.g. 963666N1), the temperature  throughout the envelope behaves as in the non-shocked case, except around minimum radius where a temperature spike occurs  in outer layers. The photosphere itself is not significantly affected (cf. Fig.\,\ref{f:layers}).

In the most extreme cases (e.g. the high opacity models 963666N3 and 963666N6 and the high $L/M$ case 064066N1), temperature and hence luminosity maxima again occur first on the surface.
In these cases, however, the shock develops in deeper layers and thence propagates outwards (Figs.\,\ref{f:opguts} and \ref{f:layers}). 
Radius minimum at the surface is substantially delayed relative to radius minimum in deeper layers. It appears to be the action of the shock which delays contraction of the outer layers and hence has the strongest influence on the phase delay between light maximum and radius minimum. 

\subsection{Phase delay}

Observations of V652\,Her clearly show a phase difference of $\approx +0.13$ cycles between maximum light and minimum radius \citep{hill81,jeffery15b}. 
A similar phase difference in the models was found by \citet{montanes02}.
Phase differences exist in classical Cepheids, but have the opposite sense (--0.25 cycles).
They have been interpreted in terms of linear non-adiabatic effects associated with the thin hydrogen ionization zone moving through mass layers almost  as a discontinuity \citep{castor68,szabo07}. 
Since the hydrogen and first helium ionization zones play no r\^ole in 
V652\,Her (too little hydrogen and too hot), it is suggested that the  second helium ionization zone will be important.
Fig.\,\ref{f:opguts} suggests that the phase delay  depends on the intensity of the shock, ranging from $\sim0.055$ in  the reference model (Fig. \,\ref{f:opguts}) to $\sim0.12$ in the most extreme example found (963666N6). 
Since shock strength appears to depend on both opacity (model 963666N6) and luminosity (064066N1), phase delay would appear to be a nonlinear function of both. 
Some insight on their separate effects is obtained from the {\bf N3} model grid (Fig.\,\ref{f:66N3} and \S\,4).

%%====================================================================

\begin{figure*}
\centering
\includegraphics[width=0.80\linewidth]{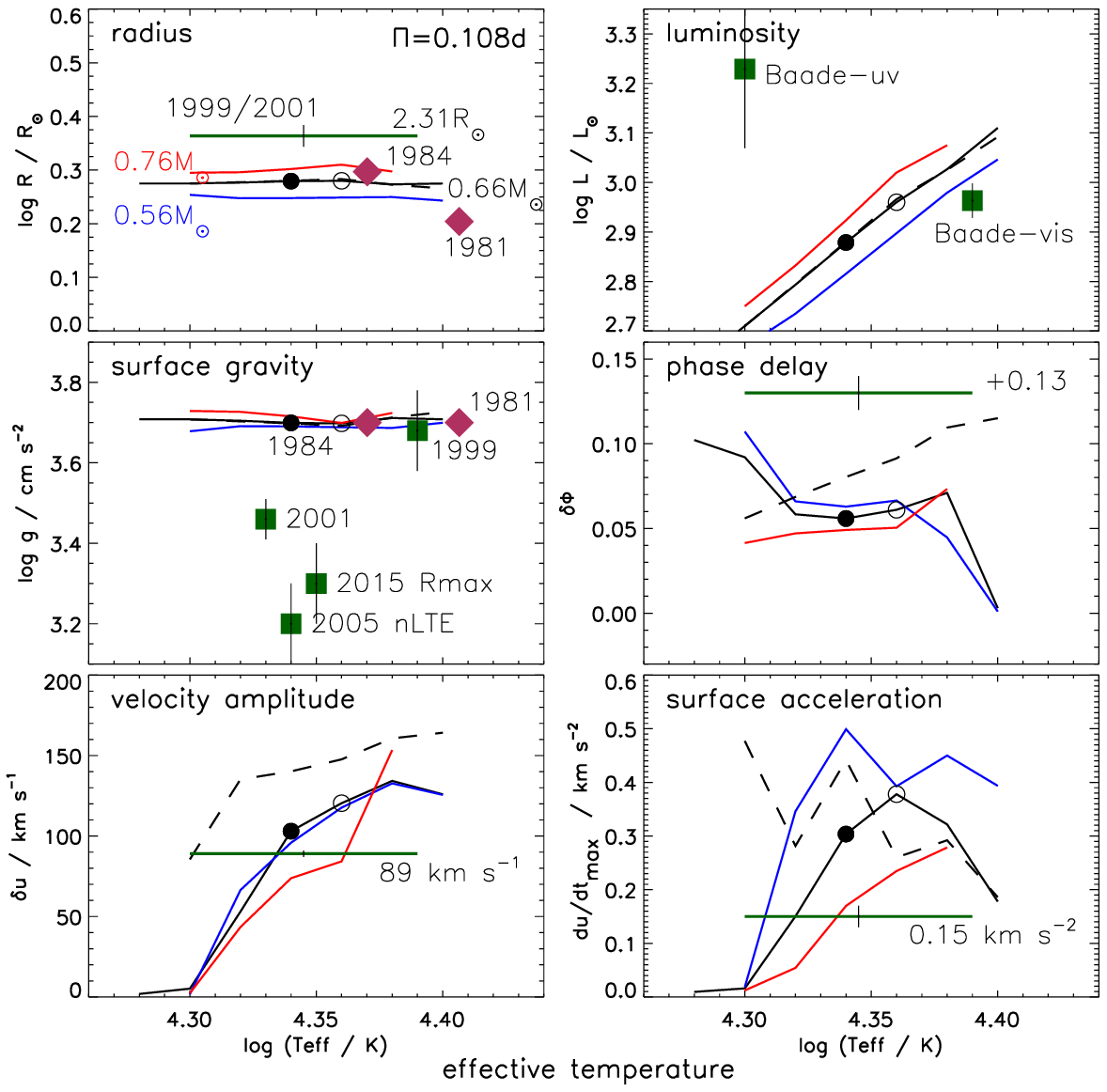}	 
 \caption{Properties of pulsation models shown in Figs.\,\ref{f:mods66} and \ref{f:mods56}-\ref{f:mods76} interpolated to a period of 0.108\,d and compared with observed properties of V652\,Her from Table\,\ref{t:pars} (thick green lines or squares). Observations from \citet{hill81} and \citet{lynasgray84} are shown as maroon diamonds
 with dates. The models are colour-coded as blue:0.56\Msolar,  black:0.66\Msolar\ and red:0.76\Msolar. A solid line indicates the reference N1 mixture, a dashed line indicates mixture {\bf N3} (iron and nickel enhanced  $\times2$).  
 An open black circle indicates the 'reference' model (963666N1), and a filled black circle suggests where further work should focus. 
} 
\label{f:obs}
\end{figure*}

\section{Comparison with V652\,Her}
\label{s:match}

The primary parameters chosen for the model computations are mass $M$, effective temperature
$T_{\rm eff}$, luminosity $L$, and chemical composition. 
The strongest observational constraint is provided by the pulsation period of 0.108\,d; this immediately constrains the radius, luminosity and surface gravity to be a slowly varying function of mass and either constant or linear in effective temperature. 
These constraints are model-independent and a direct consequence of the period mean-density relation for stars pulsating in the fundamental radial mode \citep{eddington18}. 
Figure\,\ref{f:obs} shows these quantities for the three model grids, as well as the phase delay, velocity amplitude and maximum acceleration at minimum radius, also interpolated to $\Pi=0.108$\,d.  

Several results are noteworthy. 
Some concern measurements relating to the stellar spectrum. 
Others concern properties of the models. 

The radius of V652\,Her has been measured using the \citet{baade26} method.
In simple terms this gives the radius as
\[ R = \delta R . \theta / \delta \theta,\] 
where 
\[ \delta R = \int \dot{R} {\rm d}t. \]
 $\dot{R}  \equiv {\rm d} R/{\rm d} t $ is the velocity of the stellar photosphere relative to the stellar centre and is obtained from measurements of observed and mean observed radial velocities ($v$ and $v_0$)  via a projection factor $p$ \citep{parsons72,montanes01} such that  $\dot{R} = - p (v - v_0) $.   
The angular radius $\theta$ and its variation  $\delta \theta$ are obtained from measurements of the total flux around the pulsation cycle. 
The value for the mean radius $\langle R \rangle=2.31\pm0.02\Rsolar$ cited in Table \,\ref{t:pars}  \citep{jeffery01b} appears too large by about 50\%. 
An earlier value of $1.98\pm0.21\Rsolar$ \citep{lynasgray84} is more consistent with theory.
Whether this is a consequence of assumptions in the model atmosphere analyses will be reviewed in a subsequent study.

The luminosity is derived from the distance and apparent brightness,
but may also be determined from the radius and effective temperature. Technically the two are equivalent, since the distance is related to the radius by the angular diameter, and the latter is determined by fitting the flux distribution for a given effective temperature an observed flux. 
Assuming the Baade radius above, \citet{jeffery01b} obtained contradictory distances (and hence masses and luminosities) of $0.963\pm0.006$ and $1.693\pm0.115$ kpc when normalising to visual fluxes (Baade-vis in Fig.\,\ref{f:obs}) or ultraviolet fluxes (Baade-uv), respectively. 
From Gaia Data Release 2 \citep{gaia18.dr2}, a parallax of $0.88\pm0.09$ mas led \citet{martin19.phd} to obtain a distance of $1.170^{+0.110}_{-0.130}$ kpc. In the Gaia Early Data Release 3 \citep{gaia21.dr3}, the parallax is reduced to $0.636\pm0.048$ mas. \citet{bailerjones21} consequently find a distance $1.542^{+0.114}_{-0.107}$ kpc, which is remarkably similar to the distance $1.5\pm0.1$ kpc inferred by \citet{lynasgray84}.
The pulsation period indicates that only luminosities  $\log L / \Lsolar \lesssim 3.1$ ($L < 1260 \Lsolar$) are permitted for the observed effective temperatures, which is entirely consistent with the luminosities inferred by \citet{lynasgray84} and \citet{jeffery01b} ($1072^{+341}_{-258}$ and $919\pm14 \Lsolar$, respectively). 
With sufficient precision, temperature and luminosity could be a mass discriminant.   

The surface gravity has been measured spectroscopically on a number of occasions.
Fitting the profiles of neutral helium lines using LTE model atmospheres consistently gave $\log g /\cmss \approx 3.7\pm0.2$ \citep{hill81,lynasgray84,jeffery99}.
Using higher quality data, \citet{jeffery01b} obtained the lower value $\log g /\cmss \approx 3.46\pm0.05$, averaged over the pulsation cycle.  
In an effort to resolve discrepancies in the cores of the helium lines and between different hydrogen lines, \citet{przybilla05} used a non-LTE model to find $\log g /\cmss \approx 3.2\pm0.1$. 
This was based on a sub-sample of spectra obtained around maximum radius. 
For line identification, \citet[][Appendix B]{jeffery15b} adopted a $\chi$-by-eye model with $\log g /\cmss = 3.3$ to approximate a median spectrum obtained around maximum radius. 
Pulsation arguments posit that a mean value of $\log g /\cmss \approx3.7\pm0.05$ would be consistent with the pulsation period (assuming a star in hydrostatic equilibrium). 
If $g$ is measured at $R_{\rm max}$, the effective gravity $g_{\rm eff}$ felt by the local plasma is reduced by the downward acceleration of $\Delta u / 0.8 \Pi \approx 1200\,\cmss $. 
This suggests a reduced value for 
$\log g_{\rm eff}(R_{\rm max}) /\cmss = 3.6$. 
A subsequent paper will address the question of whether adiabatic expansion following shock heating might further reduce the density of the photosphere, the physical quantity which is actually measured  by the plasma diagnostics. 

Turning to the models, several observables may be discussed in terms of model properties. 

The origin and behaviour of the phase delay between light maximum and radius minimum in the hydrodynamic models is discussed above. 
The observed value (+0.13 cycles) is larger than found in any of the models. 
A difference between visual light maximum, which is used to define the ephemeris \citep{kilkenny05},  and bolometric light maximum, which is poorly defined by current ultraviolet observations, might contribute some of the difference, and could be as much as 0.05 cycles. 
The remainder might be embedded in the treatment of the shock by the models
or in the microphysics, including the equation of state and opacity. 
Notably, while luminosity maximum at the surface precedes radius minimum, luminosity maximum at deep layers follows radius minimum by a roughly similar amount. 
The extreme non-linearity of the problem makes it difficult to quantify specific dependencies.  
For example, a minor change to the equation of state whereby the mean atomic weight in the deep interior was reduced by $\approx 0.02\%$ resulted in a reduction of the radial amplitude and an increase in the phase delay of a few per cent.

Pulsation amplitude, as measured by the peak-to-peak velocity amplitude, increases steadily with effective temperature up to a maximum around where V652\,Her lies. 
As noted earlier, this increase correlates directly with the increase in luminosity.
The observed and theoretical amplitudes match well at the effective temperature obtained by \citet{przybilla05}.
% but the corresponding  surface gravity is a factor of three too small. 
Two questions are what limits the model amplitude, and why does that limit exceed the observed amplitude over most of the permissible range of effective temperature?
One conjectures that the first is dissipation in an increasingly violent shock, and the second is that our one-dimensional model inadequately describes that dissipation. 
As a check, we reduced the viscosity parameter in the reference model by 25\% to  $C_Q=1.5$.
The phase delay $\Delta \phi$ increased from 0.062 to 0.064 cycles; the luminosity and radial-velocity amplitudes increased by a similar amount ($\approx 2.5\%$). 
Thus a significantly larger viscosity is required to reduce the theoretical amplitude,  suggesting a greater contribution from turbulence might arise in a multi-dimensional treatment. 
However, this would give the wrong sign for the phase delay. 
The alternative possibility that the observations do not sufficiently capture the amplitude of the same zones as described by the models was largely eliminated by the high-speed spectroscopy reported by \citet{jeffery15b}. 

Another way of framing the question is to ask how the pulsation properties might change with period. 
This is moot since the period of V652\,Her has decreased by  $\approx1\%$ since first measured \citep{kilkenny05}. 
Fig.\,\ref{f:obs} was generated by interpolating in the model grids for pulsation period $\Pi=0.108$\,d. By varying $\Pi$, dependency on period can be  computed easily.  
Taking into account numerical noise amongst the models and the coarseness of the grid, the phase delay is not substantially changed for a 10\% change in period, but the velocity amplitude and maximum surface acceleration are both reduced by $\approx 10 - 20\%$. 

Recognising that the mass of V652\, Her was poorly constrained by observation, the original goal of these calculations was to obtain the best possible hydrodynamic model of pulsations in V652\,Her.
Neither the effective temperature nor the radius obtained spectroscopically are fully consistent with the models. 
From the pulsation period alone, the surface gravity  $\log g / \cmss \approx 3.70\pm0.05 $, but this assumes hydrostatic equilibrium. 
Additional heating could reduce this more in line with the spectroscopic value $\approx 3.4$.  
With some latitude, our recommended initial estimate for the next phase of calculations should be $M \approx 0.66 \Msolar$, $\log \Teff / {\rm K}\approx 4.34 $, and $\log L / \Lsolar \approx 2.9$.  

Models computed around these parameters will be used as input to a time-dependent formal solution of the emergent spectrum, and will hence be applied to the exquisite observations reported by \citet{jeffery15b}. 
Together with refined distance measurement from {\it Gaia}, these should lead, in turn, to a more precise direct measurement of the radius and mass.  

A similar comparison should be carried out for the companion star BX\,Cir for which excellent data are also available \citep{kilkenny99,woolf00,woolf02,martin19.phd}. The corresponding radial velocity curve indicates no shock is present and therefore that the $L/M$ ratio may be smaller than in V652\,Her. However, with a different chemical composition containing less hydrogen and more carbon \citep{drilling98}, the current models might not be applicable; more calculations will be required.

\section{Conclusion}
\label{s:conc}

Precise observations of the unique pulsating helium star V652\,Her have provided stellar theory with a spectrum of challenges.  
This investigation set out to explore  non-linear pulsation models as a function of the primary parameters luminosity, effective temperature, and mass, as well as secondary parameters including opacity and viscosity. 
The object was to learn how these parameters might be adjusted to best-fit the observed pulsation properties, including its light and radial-velocity curves. 
 
The models demonstrate how pulsation amplitude increases with luminosity, or more generally, with luminosity-to-mass ratio.
Amplitude is less sensitive to effective temperature except towards the cool and warm boundaries of the instability region, where the amplitude is diminished.

The development of a shock during the acceleration correlates closely with amplitude and is the primary factor determining the shape of the velocity curve.  
For a given period and effective temperature, there is also an inverse correlation with mass, possibly associated with the increased inertia of the envelope. 

The origin and magnitude of the phase delay between maximum luminosity and minimum radius has proved harder to explain. 
Theoretical phase delays are generally smaller than observed  by a factor of at least 2. 
There is a moderate reduction in phase delay with increasing effective temperature, but a value at the low limit of that observed cannot resolve the problem. 
Reducing the artificial viscosity produces a small but insufficient improvement.     

It is conceivable that the chemical composition in the driving zone in V652\,Her is richer in iron and nickel than observed on the surface.
\citet{byrne20} showed that radiative levitation  enriches the iron-bump driving zone in models of contracting low-mass pre-white dwarfs at $\Teff \gtrsim 20\,000$K, and earlier for more massive stars, sufficiently to trigger pulsations in stars observed as blue large amplitude pulsators \citep[BLAPs:][] {pietrukowicz17,kupfer19}.
Whether such enrichment also occurs in post-white dwarf merger stars has yet to be tested.
Meanwhile it will be interesting to extend the non-linear calculations presented here to the case of the BLAPs in the hope that high-quality observations will in future demonstrate the ubiquity and nature of shocks amongst these stars.

\section*{Acknowledgements}

CSJ acknowledges support from the UK Science and Technology Facilities Council (STFC) Grant No. 
ST/M000834/1. 
PMR acknowledges financial support from the Spanish Ministry of Economy and Competitiveness (MINECO) under the 2011 Severo Ochoa Program MINECO SEV-2011-0187.
The Armagh Observatory and Planetarium is funded by direct grant form the Northern Ireland Dept for Communities.

\section*{Data Availability}
The summary files and internal structures for the last 240,000 timesteps of each model are stored as binary files. They will be provided on reasonable request to the authors. The surface variation of velocity, luminosity and radius as a function of phase over 1.4 pulsation cycles are available as ASCII files as supplementary online material. 

\bibliographystyle{mnras}
\bibliography{ehe.bib}

\appendix
\renewcommand\thefigure{A.\arabic{figure}} 
\renewcommand\thetable{A.\arabic{table}} 

\section{Additional figures}
\begin{figure*}
\centering
\includegraphics[width=1.0\textheight,angle=90]{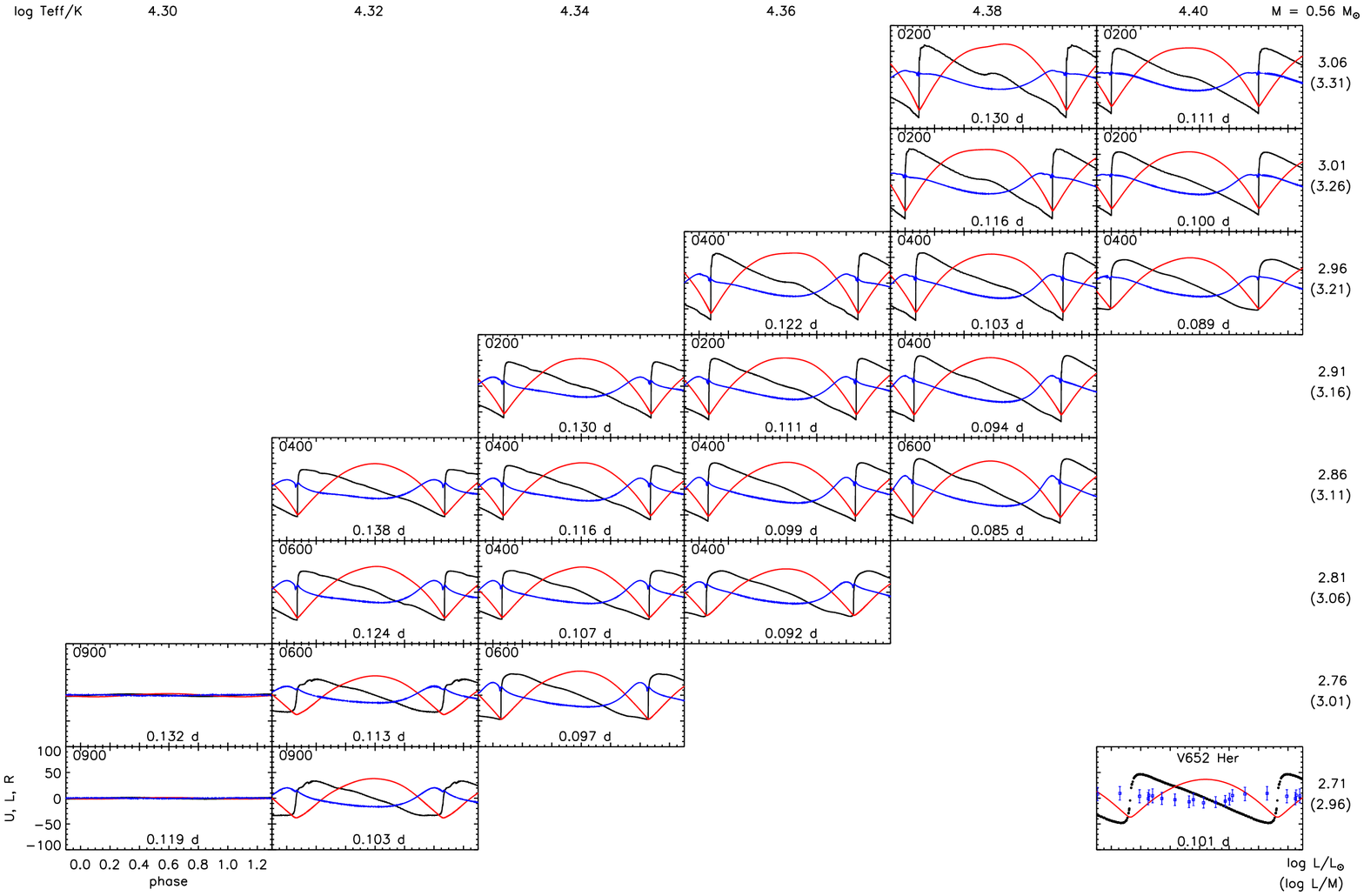}
\vspace*{-12mm}
\caption{As Fig.\,\ref{f:mods66} for $M/\Msolar=0.56$. }
\label{f:mods56}
\end{figure*}

\begin{figure*}
\centering
\includegraphics[width=1.0\textheight,angle=90]{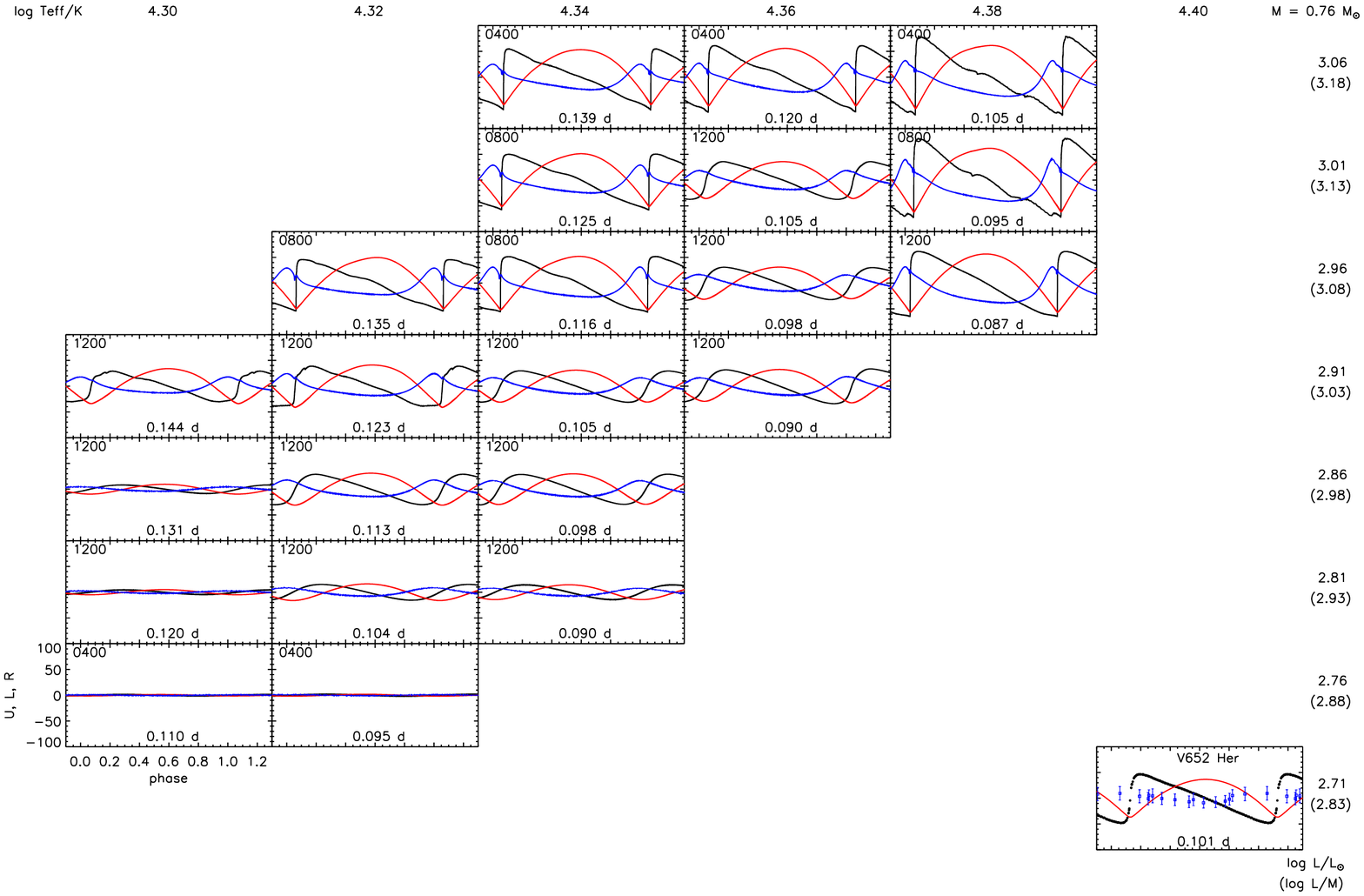}
\vspace*{-12mm}
\caption{As Fig.\,\ref{f:mods66} for $M/\Msolar=0.76$.}
\label{f:mods76}
\end{figure*}

\begin{figure*}
\centering
\includegraphics[width=1.0\textheight,angle=90]{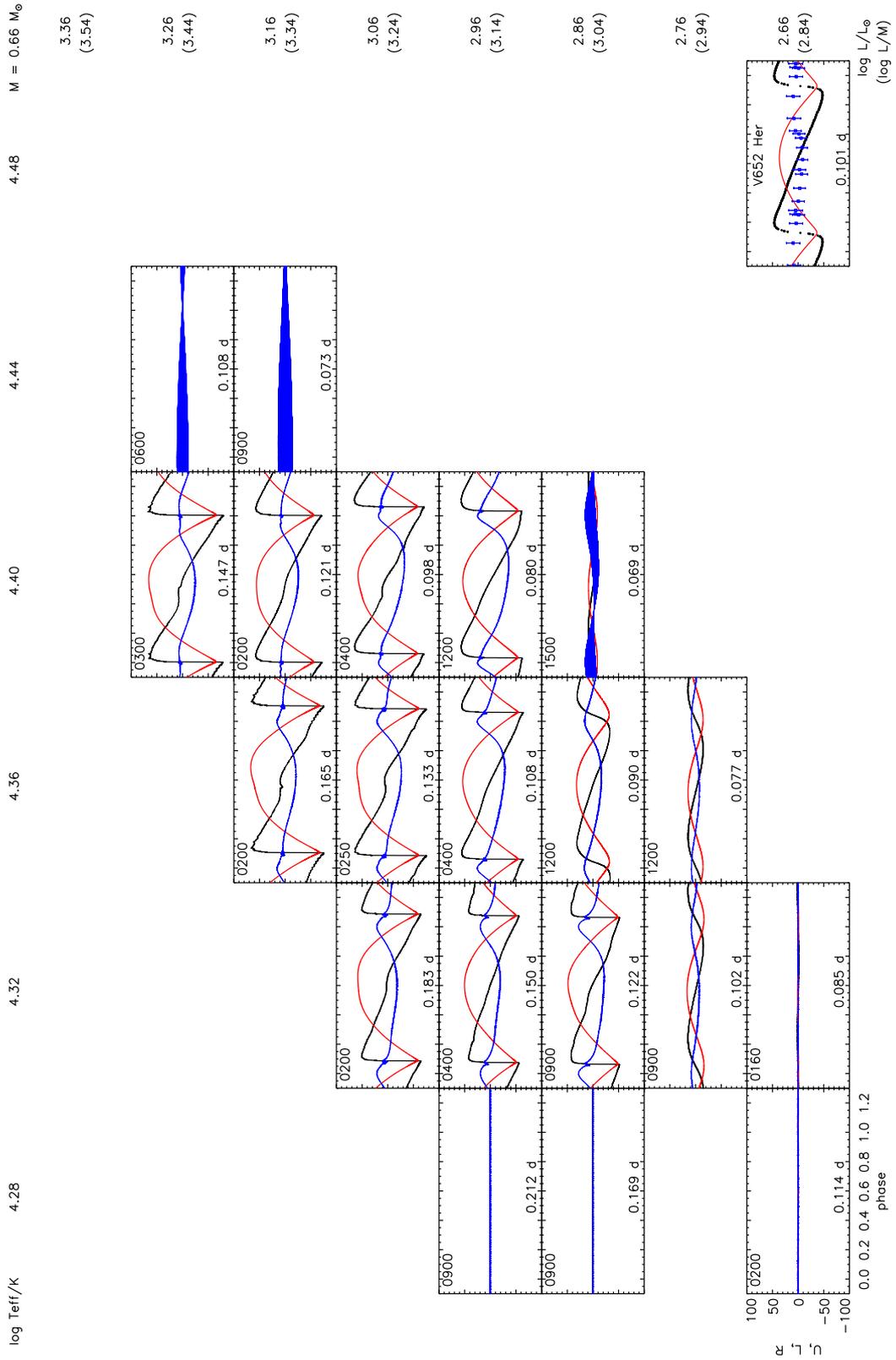}
\vspace*{-12mm}
\caption{As Fig.\,\ref{f:mods66} for $M/\Msolar=0.66$ showing extensions to lower and higher luminosity and effective temperature. }
\label{f:mods66c}
\end{figure*}

This appendix includes Figs.\,\ref{f:mods56}-\ref{f:mods66c} which are equivalent to Fig.\,\ref{f:mods66} for additional masses and mixtures.

\begin{figure*}
\centering
\includegraphics[width=1.0\textheight,angle=90]{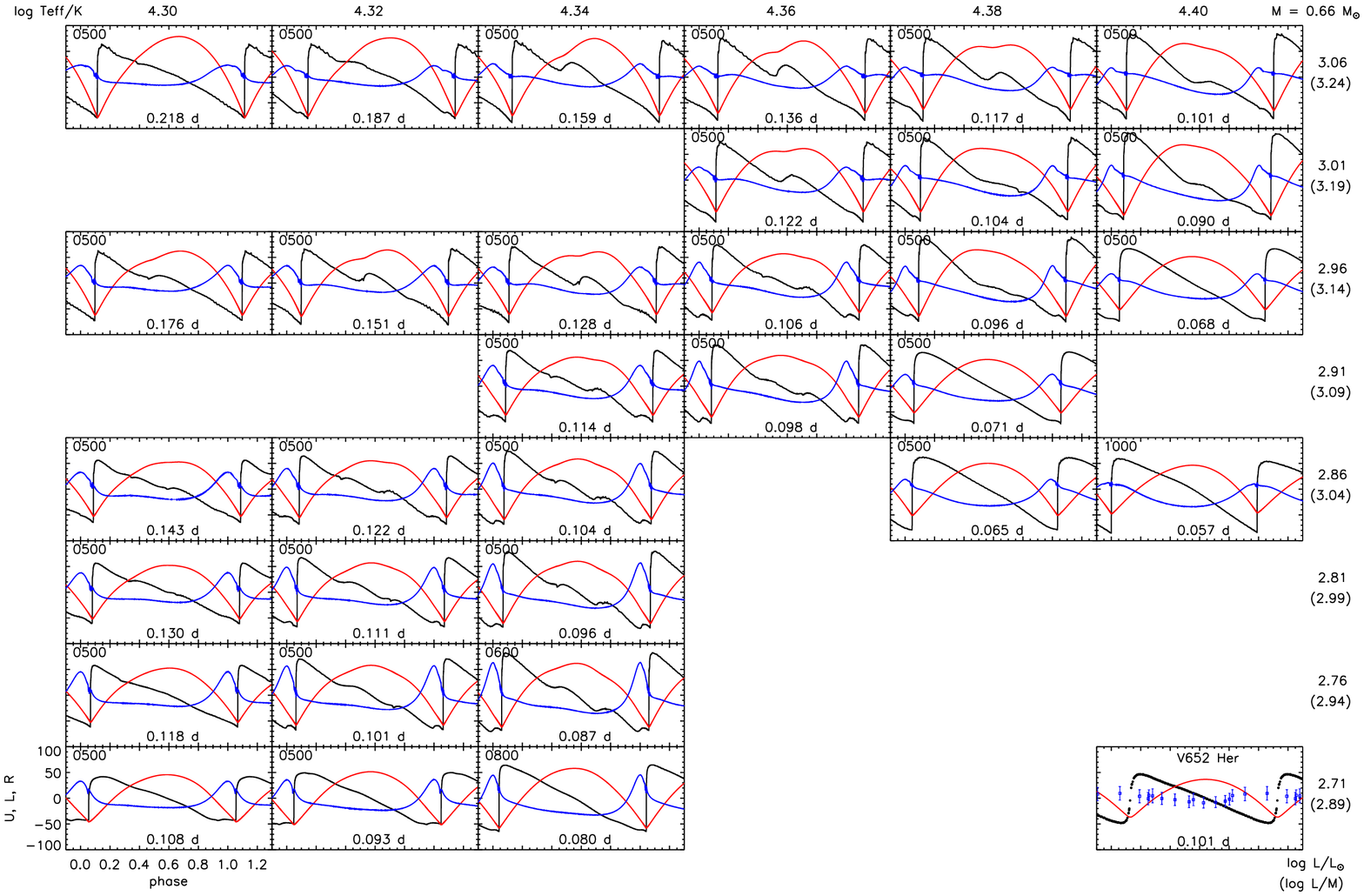}
\vspace*{-12mm}
\caption{As Fig.\,\ref{f:mods66} for mixture {\bf N3} (iron and nickel abundances enhanced $\times 2$). }
\label{f:66N3}
\end{figure*}

\renewcommand\thefigure{B.\arabic{figure}} 
\renewcommand\thetable{B.\arabic{table}} 
\section{Supplementary data}
Table\,\ref{t:supp} shows a list of files containing supplementary material containing the data shown in  Fig.\,\ref{f:mods66} and  Figs.\,\ref{f:mods56}-\ref{f:mods66c} as well as IDL procedures for reading reading and plotting specified models in the form of Fig.\,\ref{f:96636N1}.  
Table\,\ref{t:surface} shows a file fragment from one of the data files. 
The IDL procedures are self-documented with instructions for use.

\begin{table*}
    \caption{Files containing Supplementary Data available online.}
    \centering
    \begin{tabular}{ll}
    File & Description \\
    \verb| surf_read.pro | & IDL procedure to read specified pulsation model \\
    \verb| surf_plot.pro | & IDL procedure to plot specified pulsation model \\
    \verb| surf_m56.log  | & Surface properties for pulsation models at 0.56 $\Msolar$, mix {\bf N1} \\
    \verb| surf_m66.log  | & Surface properties for pulsation models at 0.66 $\Msolar$, mix {\bf N1}  \\
    \verb| surf_m76.log  | & Surface properties for pulsation models at 0.76 $\Msolar$, mix {\bf N1}  \\
    \verb| surf_m66N3.log  | & Surface properties for pulsation models at 0.66 $\Msolar$, mix {\bf N3}  \\
    \end{tabular}
    \label{t:supp}
\end{table*}

\begin{table*}
\caption{Fragment from one of the supplementary online files containing data used in the construction of Figs. \ref{f:mods66} and Figs.\,\ref{f:mods56}-\ref{f:mods66c} and describing the variation of velocity in \kmsec, radius (per cent) and luminosity (per cent) with pulsation phase. Each data file contains all models for a given mass and mixture; each model is demarcated by a header line containing its label and other parameters, as shown. An IDL procedure to read these files is provided with the online material. }
\begin{tabular}{l}
\verb| MODEL  log10(<L>/Lsun)  log10(<Teff>/K)  M/Msun  log10(<R>/Rsun)  P/d  | \\
\verb| phase  RV/(km/s)  100*(R-<R>)/<R>  100*(L-<L>)/<L> |\\
 \\
\verb| 963666N1  2.958 4.360 0.660 0.280 0.1083 | \\
\verb| -0.1000 -43.37   0.654  -2.463 | \\
\verb| -0.0975 -44.02   0.575  -1.627 | \\
\verb| -0.0953 -44.15   0.512  -0.990 | \\
\verb| ... | \\
\end{tabular}
\label{t:surface}
\end{table*}

\label{lastpage}
\end{document}